\documentclass[fleqn,usenatbib]{mnras}

\usepackage{graphicx}
\usepackage{longtable}
\usepackage{supertabular}
\usepackage{rotating}
\usepackage{amsmath,amssymb}
\usepackage{color}
\usepackage{multirow}
\usepackage{arydshln}

\newcommand{\gtsim}{\protect\raisebox{-0.5ex}{$\:\stackrel{\textstyle >}{\sim}\:$}}
\newcommand{\ltsimscript}{\protect\raisebox{-0.5ex}{$\stackrel{\scriptstyle <}{\sim}$} }

\newcommand{\suzaku}{{\it Suzaku} }
\newcommand{\nustar}{{\it NuSTAR}}

\title[Application of X-ray reflection model to V1223 Sgr]{Application of a new X-ray reflection model to V1223 Sagittarii}

\author[T. Hayashi, T. Kitaguchi and M.Ishida]{Takayuki Hayashi$^{1, 2}$
\thanks{E-mail: thayashi@umbc.edu} 
Takao Kitaguchi$^{3, 4}$
and Manabu Ishida$^{5, 6}$
\\
$^{1}$NASA, Goddard Space Flight Center, Code 662, Greenbelt, MD20771, USA\\
$^{2}$Department of Physics, University of Maryland Baltimore County, 1000 Hilltop Circle 
Baltimore, Maryland 21250, USA\\
$^{3}$ RIKEN Cluster for Pioneering Research, 2-1 Hirosawa, Wako, Saitama 351-0198, Japan\\
$^{4}$ RIKEN Nishina Center, 2-1 Hirosawa, Wako, Saitama 351-0198, Japan\\ 
$^{5}$The Institute of Space and Astronautical Science/JAXA, 3-1-1 Yoshinodai, Chuo-ku, Sagamihara 252-5210, Japan\\\
$^{6}$Department of Physics, Tokyo Metropolitan University, 1-1 Minami-Osawa, Hachioji, Tokyo 192-0397}
\begin{document}

\date{}

\pagerange{\pageref{firstpage}--\pageref{lastpage}} \pubyear{2012}

\maketitle

\label{firstpage}

\begin{abstract}

In intermediate polars (IPs), the intrinsic thermal emissions from white dwarfs (WDs) have typically been studied. Few reports have analyzed X-ray reflections from WDs. We recently developed an elaborate IP-reflection spectral model. Herein, we report the first application of a reflection model 
with an IP thermal model to the spectra of the brightest typical IP V1223\, Sagittarii observed by the {\it Suzaku} and {\it NuSTAR} satellites.
The model reasonably reproduces the spectra within the range of 5--78\,keV
and estimates the WD mass as 0.92$\pm$0.02\, $M_{\odot}$.
The WD mass estimated by the proposed model is consistent with that measured using an active galactic nuclei reflection model and a partial covering absorption model. However, the choice of incorrect parameter values, such as an unsuitable fitting energy band and an incorrect metal abundance, was found to introduce 
systematic errors 
(e.g., $\ltsimscript$\, 0.2\, $M_{\odot}$ in the WD mass)
in the WD mass measurement.
Our spin phase-resolved analysis resulted in discoveries regarding the
modulations of the equivalent width of the fluorescent iron K$_\alpha$ line 
and the angle between the post-shock accretion column and the line-of-sight (viewing angle). 
The viewing angle anti-correlates approximately with the X-ray flux and 
has average and semi-amplitude values of 55$^\circ$ and 7$^\circ$, respectively, 
which points toward two WD spin axis angles from the line-of-sight of 55$^\circ$ and 7$^\circ$, respectively. 
Both estimated spin axis angles are different from the reported system inclination of 24$^\circ$.

\end{abstract}


\section{Introduction}\label{sec:intro}

Cataclysmic variables (CVs) are close binary systems
consisting of a companion late-type main-sequence star and a white dwarf (WD).
The late-type star fills the Roche lobe and 
feeds its gaseous material to the WD through the inner Lagrange point.
CVs comprising a highly magnetized WD ($B > 0.1$~MG) are called magnetic CVs (mCVs), which are categorized as polar or intermediate polar (IP) 
systems. In polar systems, the magnetic field is quite
strong ($B > 10$\,MG) and can channel the gaseous flow directly from the companion star.
In IPs, the magnetic field is moderate ($0.1 < B < 10$\,MG), and an accretion disk forms around the WD.
The magnetic field tears the gas from the accretion disk near the Alfv$\acute{\rm e}$n radius and channels it. The channeled gas
falls toward the WD at an almost free-fall velocity and is accelerated up to hypersonic speeds. A strong shock is formed close to the WD surface,
heats the gas, and generates highly ionized plasma.
In IP systems, the plasma flow is cooled by an X-ray thermal emission, and the plasma flow then settles onto the WD surface, 
which is called the post-shock accretion column (PSAC).
The PSAC irradiates the WD with X-rays, and 
the WD shines in the X-ray owing to the reflection.

The X-ray observation enables us to measure the WD mass independent of 
the dynamical measurement and to determine the PSAC geometry.
The WD model is still controversial, particularly under extreme conditions (for example, \citealt{2004A&A...419..623Y,2013ApJ...767L..14D}),
and the WD mass is a fundamental physical parameter used to constrain the model.
However, the well-established dynamic WD mass measurement is difficult, except in systems showing an eclipse 
because of the uncertainty in the orbital inclination angle.
By contrast, with X-ray spectroscopy, the WD mass can be measured by measuring the plasma temperature, which is correlated with the gravitational potential.
Moreover, the reflection enables us to measure the height of the PSAC and the angle between the PSAC and the line-of-sight.
Height is a fundamental parameter that specifies the PSAC and constrains the WD mass.
The PSAC angle is modulated by the WD spin, which allows us to estimate the spin axis angle.
The spin axis angle provides the direction of the angular momentum,
which is new information regarding the dynamics of a binary system.

The thermal X-ray spectrum of PSAC has been well studied.
The PSAC has been hydrodynamically modeled (\citealt{1973PThPh..49..776H}, \citealt{1973PThPh..49.1184A}, 
\citealt{1983ApJ...268..291I}, \citealt{1994ApJ...426..664W}, \citealt{1996A&A...306..232W}, \citealt{1998MNRAS.293..222C}
\citealt{1999MNRAS.306..684C}, \citealt{2005A&A...440..185C},
\citealt{2005MNRAS.360.1091S}, \citealt{2007MNRAS.379..779S} and \citealt{2014MNRAS.438.2267H}).
Such studies
involved various physical effects:\,gravitational potential release,
cross-section 
convergence, and variation in the accretion rate per unit area (called the ``specific accretion rate''). PSAC models have been employed to measure the WD mass of the mCVs:
(\citealt{1991ApJ...367..270I},
\citealt{1997ApJ...474..774F},
\citealt{1998MNRAS.293..222C}
\citealt{1999MNRAS.306..684C},
\citealt{1999ApJS..120..277E},
\citealt{2000MNRAS.316..225R},
\citealt{2005A&A...435..191S},
\citealt{2009A&A...496..121B},
\citealt{2009MNRAS.392..630L},
\citealt{2010A&A...520A..25Y} and
\citealt{2014MNRAS.441.3718H}).

Less attention has been paid to the X-ray reflection from the WD surface
despite 
its prominent spectral features. 
The reflection consists of thermal X-rays escaping from the WD through scattering and/or fluorescence. 
The reflection spectrum has distinctive features:\,fluorescent iron K$_\alpha$ lines, 
a Compton hump at approximately 10--30keV, and Compton shoulders following the emission lines.
These features enable us to study the geometry between a PSAC and a WD. However, these features can be a nuisance when obtaining thermal plasma parameters unless they are correctly modeled. 

Several studies have considered the reflection in their spectral analysis 
(e.g., \citealt{1995MNRAS.272..749B}, \citealt{1998MNRAS.293..222C}, 
\citealt{2000MNRAS.315..307B}, \citealt{2011PASJ...63S.739H}, \citealt{2015ApJ...807L..30M}, and \citealt{2018MNRAS.476..554S}),
invoking reflection models developed for 
active galactic nuclei (AGN) (e.g.,\,\citealt{1995MNRAS.273..837M} and \citealt{1991MNRAS.249..352G})
or for the mCV \citep{1996A&A...312..186V}.
However, these models do not incorporate the stratified structure of the PSAC, finite height of the PSAC, or the WD curvature. 
Moreover, the features incurred by the scattering and fluorescence were handled separately. 

Recently, we developed an IP reflection model using a Monte Carlo simulation \citep{2018MNRAS.474.1810H}.
In the model, a finite-length columnar source on a spherical reflector irradiated the reflector 
with X-rays, the spectrum of which was determined by 
the PSAC stratified structure calculated by \cite{2014MNRAS.438.2267H}.
The reflector was cold and neutral,
and its radius was determined based on the WD mass \citep{1972ApJ...175..417N}.
The simulation involved X-ray interactions with atoms, i.e.,
coherent/incoherent scattering and photoelectric absorption.
The photoelectric absorption by iron and nickel
is followed by K$_{\alpha1, 2}$ and K$_\beta$ fluorescent emissions with the corresponding fluorescent yield.
The reflection model has five fitting parameters:\,the WD mass ($M_{\rm WD}$), 
specific accretion rate ($a$), elemental abundance ($Z$), 
the angle between the PSAC and line of sight ($i$), and normalization.
These parameters are also common to the IP thermal model of \cite{2014MNRAS.438.2267H}.
 
We selected V1223\,Sagittarii (V1223\,Sgr) as the first target to 
demonstrate how well the IP reflection model reproduces the observed spectrum
and the parameters that can be measured. 
V1223\,Sgr is a typical IP and one of the brightest.
The other brightest IPs, EX\,Hydrae and V2400\,Ophiuchus, are
somewhat extraordinary owing to their extremely low X-ray luminosity \citep{2005A&A...435..191S} 
and the discless feature \citep{1995MNRAS.275.1028B}, respectively.
V1223\,Sgr was observed for 190 and 20\,ks using the {\it Suzaku} and nuclear spectroscopic telescope array ({\it NuSTAR}) satellites,
respectively (\S\ref{sec:obs}), the data of which are publicly available.
The {\it Suzaku} data have a large effective area for fluorescent iron K$\alpha$ lines,
and the fluorescent line was resolved from the thermal K$_{\alpha}$ lines. 
The {\it NuSTAR} data have a 
high signal-to-noise ratio in the Compton hump energy band.
 
In this paper, we present an application of the IP reflection model to the
V1223\,Sgr data acquired by the {\suzaku} and 
{\nustar}) satellites. 
We describe the observations and data reduction in \S\ref{sec:obs} 
and the model application in \S\ref{sec:app}.
We discuss the reflection spectral modeling, the effects of various parameters on the WD mass estimation, and the geometry in \S\ref{sec:dis}.
The conclusions are presented in \S\ref{sec:con}.

\section{Observations and Data reduction}\label{sec:obs}

We used 
{\it Suzaku} \citep{2007PASJ...59S...1M} and {\it NuSTAR} \citep{2013ApJ...770..103H} archival data for V1223\,Sgr. 
Table\,\ref{table:obs} shows a summary of the observation.
We applied a barycentric correction to each 
dataset 
\citep{2008PASJ...60S..25T}.

\subsection{\it Suzaku}
{\it Suzaku} has two instruments: \, an X-ray imaging spectrometer (XIS; \citealt{2007PASJ...59S..23K}) 
and a hard X-ray detector (HXD; \citealt{2007PASJ...59S..53K}; \citealt{2007PASJ...59S..35T}).
The XIS has an imaging capability with the aid of an X-Ray Telescope (XRT; \citealt{2007PASJ...59S...9S}).
The 2007 and 2014 observations (hereafter referred to as 07S and 14S) were conducted at the nominal positions of the HXD 
and XIS, respectively.
The exposure time is tabulated in Table\,\ref{table:obs}.
The 07S and 14S data were screened through 
the {\it Suzaku} processing pipeline version 3.0.22.43 and 3.0.22.44, respectively,
with the latest calibration database (20160607 for XIS, 20110630 for XRT, and 20110913 for HXD).

\begin{table*}
\centering
 \caption{Observation summary of V1223\,Sgr by \suzaku and \nustar.}\label{table:obs}
 \begin{tabular}{cccccc}
 \hline
 Observation ID & Observation date (UT) & data set name & Aim point & Detector & Exposure (ks)$^{\ast}$\\
 \hline
\multicolumn{5}{c}{\suzaku}\\
 402002010 & 2007 April 13--14th & 07S & HXD & XIS & 60.7\\
& & & & HXD & 46.3\\
\hdashline
 408019010 & 2014 March 29--30th & \multirow{4}{*}{14S} & XIS & XIS & 29.4\\
& & & & HXD & 26.1\\
 408019020 & 2014 April 10--14th &  & XIS & XIS & 150.8\\
& & & & HXD & 146.3\\
 \hline
 \multicolumn{5}{c}{\nustar}\\
30001144002 & 2014 September 16 & 14N& - & FPMA/FPMB&20.4\\
  \hline
$^{\ast}$ Exposure time after data screening.
  \end{tabular}
\end{table*}

\subsubsection{XIS}
We extracted the source events from a circle of radius 4$\acute{.}$34 around the image center.
The background events fall in the annulus between radii of 4$\acute{.}$34 and 8$\acute{.}$68,
excluding those regions irradiated by the $^{55}$Fe calibration sources
and the detector edges.
One of the XIS sensors XIS0 in 14S has an unusable area owing to an anomaly 
that 
was also excluded from the background region.
The spectra of the two front-illuminated (FI) sensors (XIS0 and 3) were combined. 
During all of the observations, the other FI sensor XIS2 was deemed
unusable because of an anomaly that occurred in November 2006.
Subsequently, the data acquired during the two periods in 2014 were combined. 

\subsubsection{HXD-PIN}
We used the ``tuned'' non-X-ray background (NXB) spectrum version 2.0
provided by the instrument team \citep{2009PASJ...61S..17F} for the 07S HXD-PIN data.
The cosmic X-ray background (CXB) spectrum was generated by convolving the 
spectral model of \cite{1987PhR...146..215B} 
with the detector response.
The NXB and CXB spectra were subtracted from the HXD-PIN data. 
We did not use the S14 HXD-PIN data because only a ``quick'' 
NXB spectrum was provided, which had larger systematic errors.

\subsection{\it NuSTAR}

{\nustar} has two co-aligned hard X-ray telescopes. 
Each telescope focuses celestial X-rays onto
a focal plane module A or B (FPMA or FPMB). 
The observation of V1223\,Sgr was conducted in September 2014 (hereafter referred to as 14N).
We reprocessed the data using {\it NuSTAR} data analysis software ({\tt NuSTARDUS} v.1.7.1) 
and the calibration files of 20171002 to extract the cleaned events.
The source events were extracted within a circle of 2$\acute{.}$5 
radii around the image center, 
whereas the background events were extracted 
within an annulus between radii of 2$\acute{.}$5 and 5$\acute{.}$0 by excluding the detector edges.

\section{Spectral model fitting}\label{sec:app}

The IP thermal model of \cite{2014MNRAS.438.2267H}
and the reflection model of \cite{2018MNRAS.474.1810H}
(hereafter referred to as {\sc acrad$_{\rm th}$} and {\sc acrad$_{\rm ref}$}, respectively) 
have been compiled into table models that are available on the XSPEC \citep{1996ASPC..101...17A}.
We fitted the IP models to the V1223\,Sgr spectrum using {\sc XSPEC} version 12.9.1,
binding the same parameters in both models.
For comparison, we also fitted 
an AGN reflection model ({\sc reflect} in {\sc XSPEC}; \citealt{1995MNRAS.273..837M})
or a partial covering absorption (PCA) model ({\sc pcfabs} in {\sc XSPEC})
instead of {\sc acrad$_{\rm ref}$}.
Phenomenologically, the PCA model may reproduce 
the spectral shape of the Compton hump \citep{1998MNRAS.293..222C}.
With {\sc reflect} and {\sc pcfabs}, a narrow Gaussian model was added to the fluorescent iron K$_\alpha$ line,
whose central energy and width were fixed at 6.4 and 0\,keV, respectively.
The abundance of iron and other elements in {\sc reflect}
were bound to that of {\sc acrad$_{\rm th}$}.
For all cases, the photoelectric absorption was considered using 
{\sc phabs}.
In summary, we used the three models of 
 {\sc phabs$\times$(acrad$_{\rm th}$+acrad$_{\rm ref}$)},  {\sc phabs$\times$(reflecting $\times$acrad$_{\rm th}$+Gaussian)}, and 
 {\sc phabs$\times$(pcfabs$\times$acrad$_{\rm th}$+Gaussian)}, which are
hereafter referred to as {\tt IP-reflect}, {\tt AGN-reflect}, and {\tt PCA-reflect}, respectively.
We excluded data below 5\,keV 
to avoid the complicated absorption feature
created by the multicolumnar absorber
\citep{1999ApJS..120..277E}.
The exclusion of $<5$\,keV data is reasonable 
because the main features of the maximum plasma temperature 
depending on the WD mass and the reflection appear at above 5 keV.

\subsection{Spin averaged spectrum}\label{sec:ave_spe_fit}

\begin{figure}
\includegraphics[width=80mm]{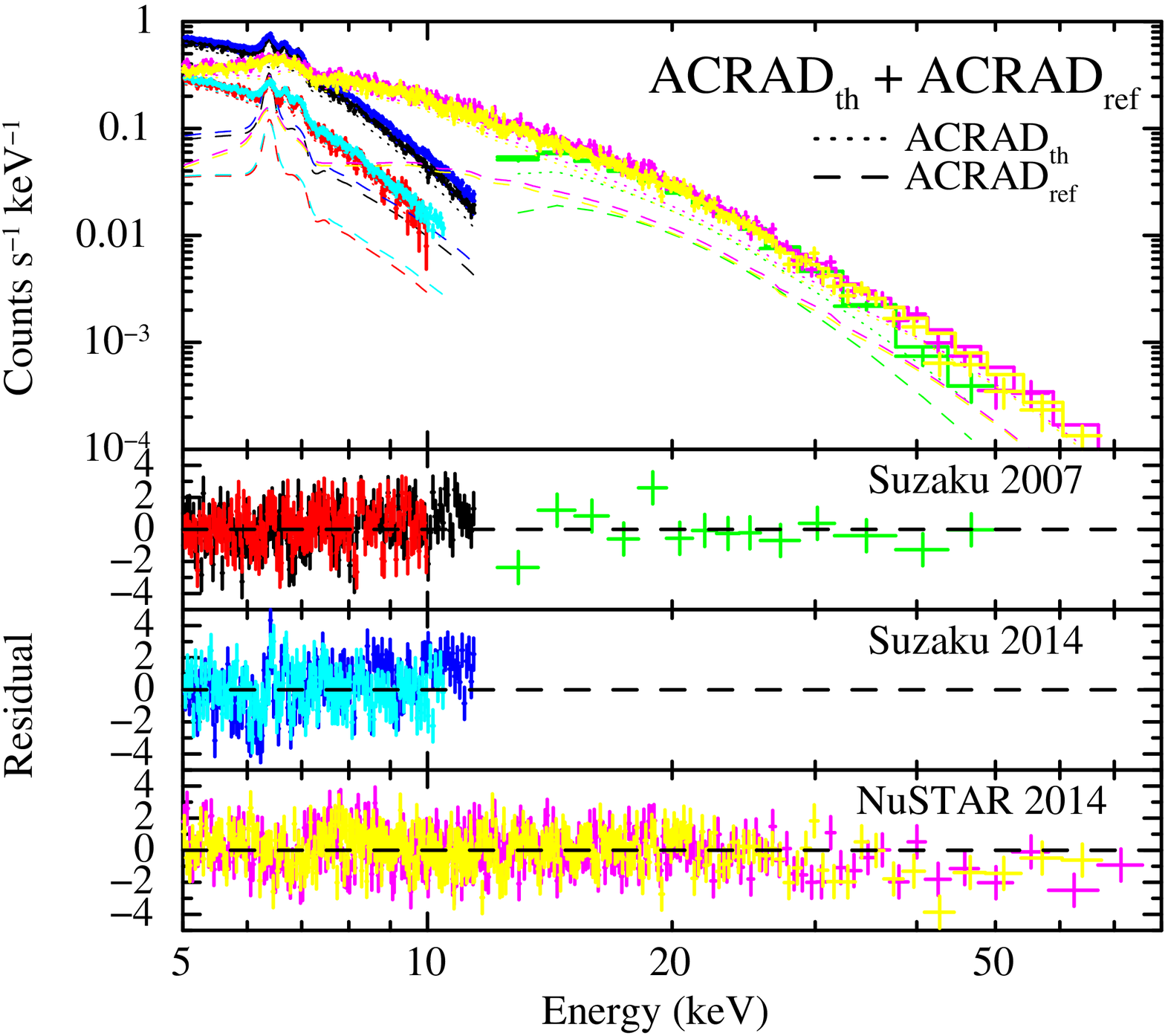}
\includegraphics[width=80mm]{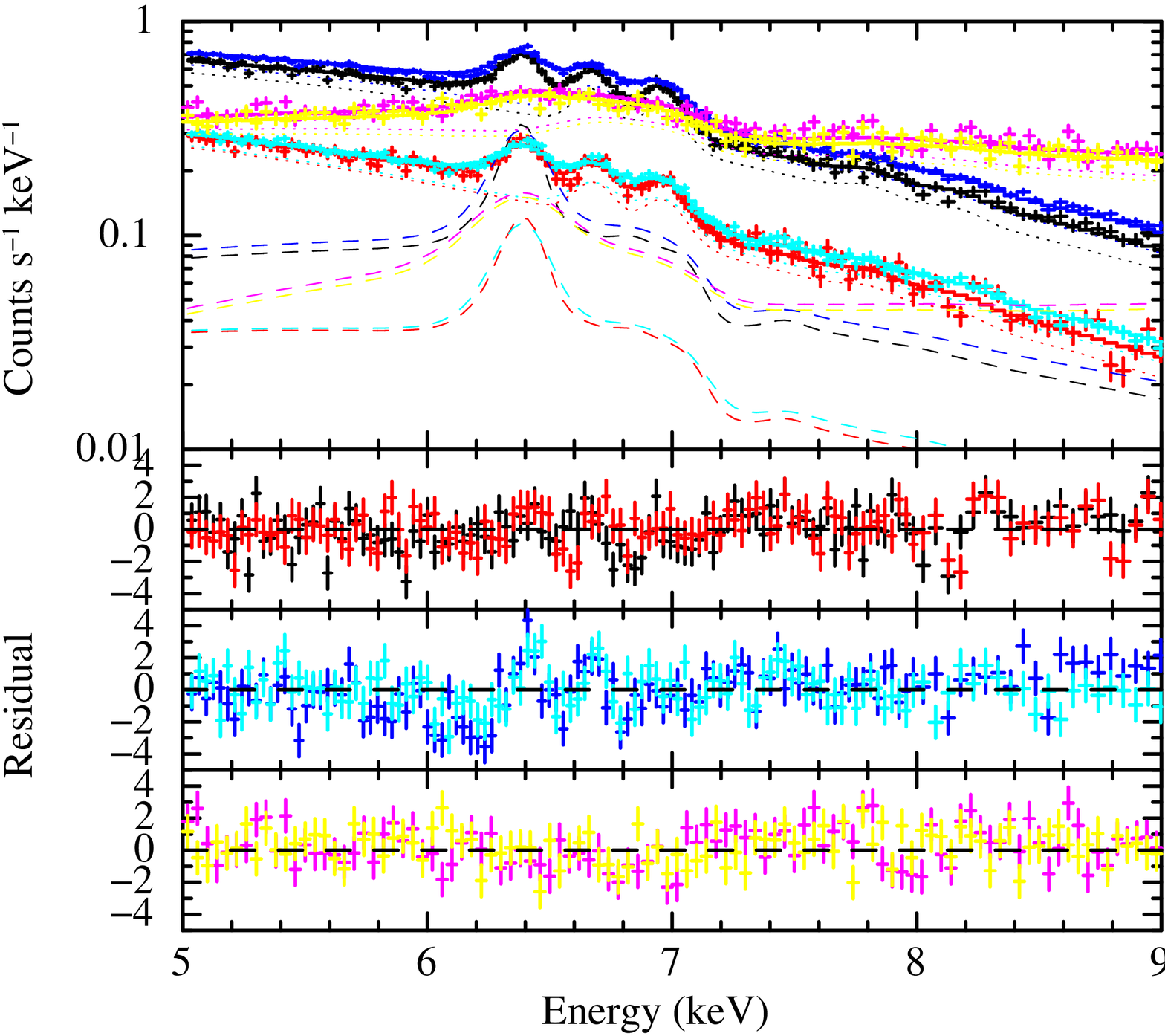}
 \caption{Top: 07S-FI (black), 07S-BI (red), 07S-PIN (green), 
 14S-FI (blue), 14S-BI (light blue), 14N-FPMA (magenta), and 14N-FPMB (yellow) spectra
 with the best-fit {\tt IP-reflect} 
 model.
The three lower panels show the residuals of 07S, 14S, and 14N from the top 
in units of $\sigma$.
The dotted and dashed lines show the thermal ({\sc acrad$_{\rm th}$}) and reflection ({\sc acrad$_{\rm ref}$}) components, respectively.
Bottom: Blowup of top panel between 5 and 9\,keV.}
\label{fig:ave_fit_acrad}
\end{figure}

\begin{figure}
\includegraphics[width=80mm]{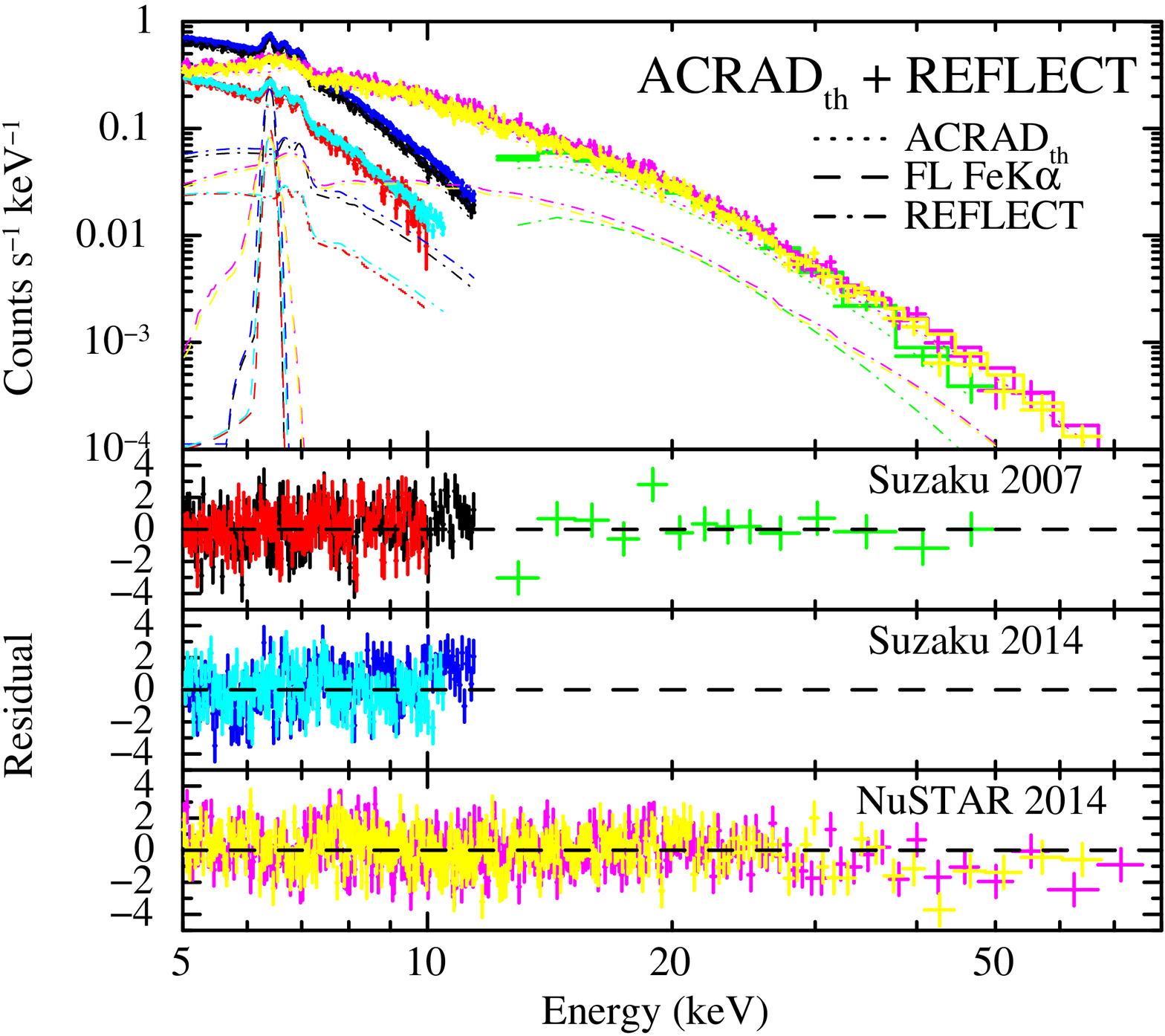}
\includegraphics[width=80mm]{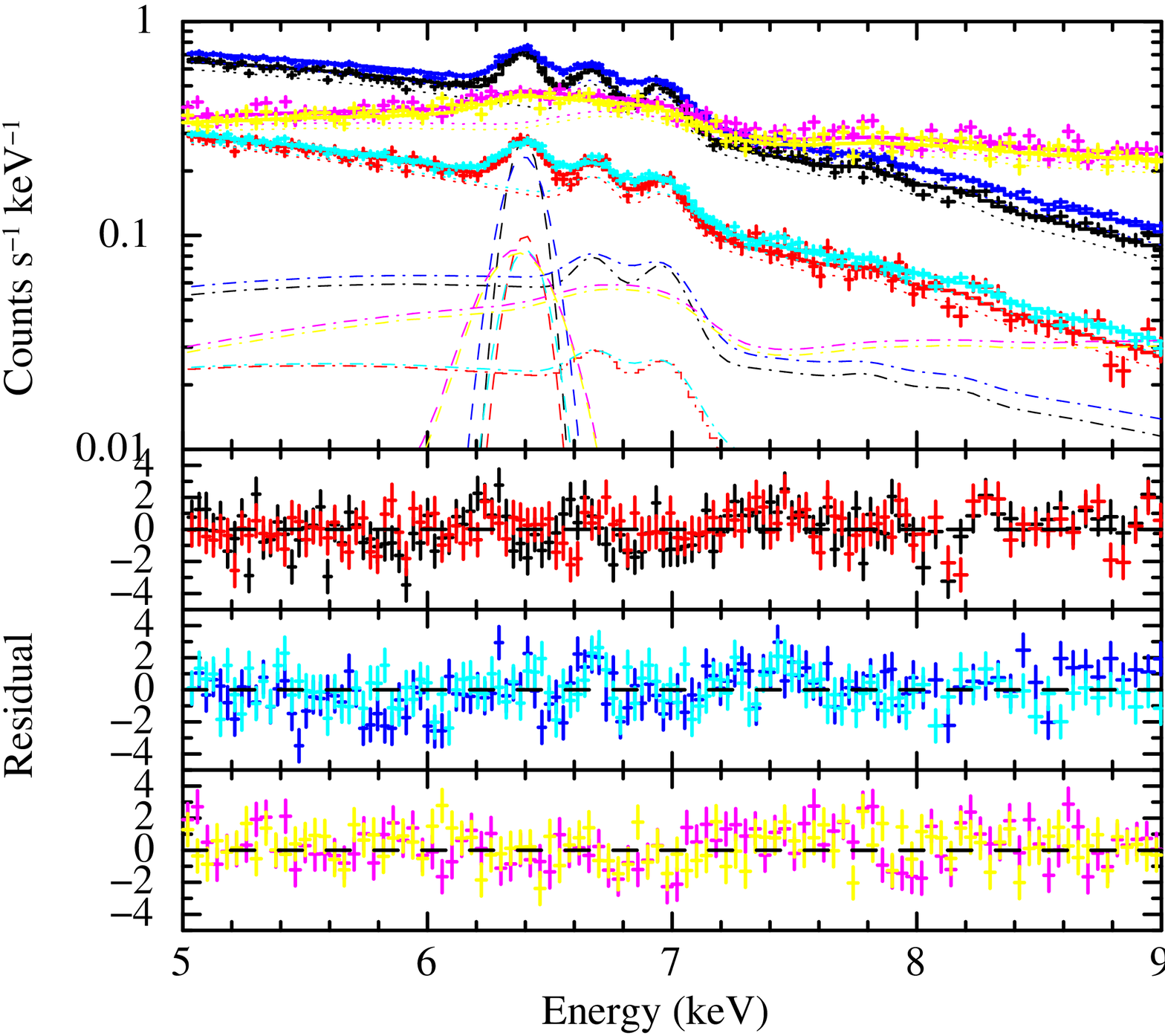}
 \caption{Same as Figure\,\ref{fig:ave_fit_acrad}, although the fitting model is 
{\tt AGN-reflect}.
 The dotted, dotted-dashed, and dashed lines show the thermal ({\sc acrad$_{\rm th}$}), fluorescent iron K$_\alpha$ line ({\sc Gaussian}), and reflection ({\sc reflect}) components, respectively.
}
\label{fig:ave_fit_reflect}
\end{figure}

\begin{figure}
\includegraphics[width=80mm]{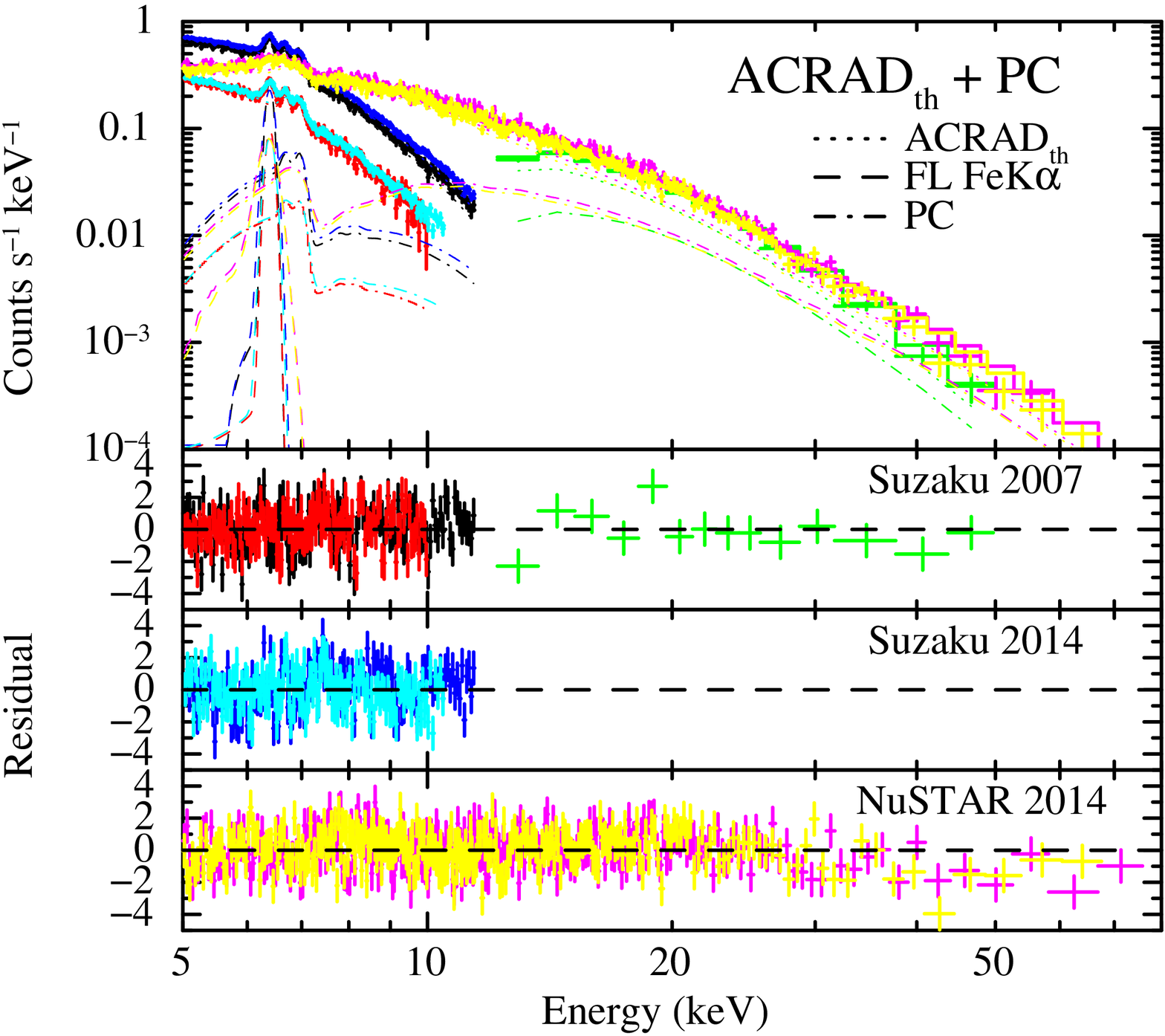}
\includegraphics[width=80mm]{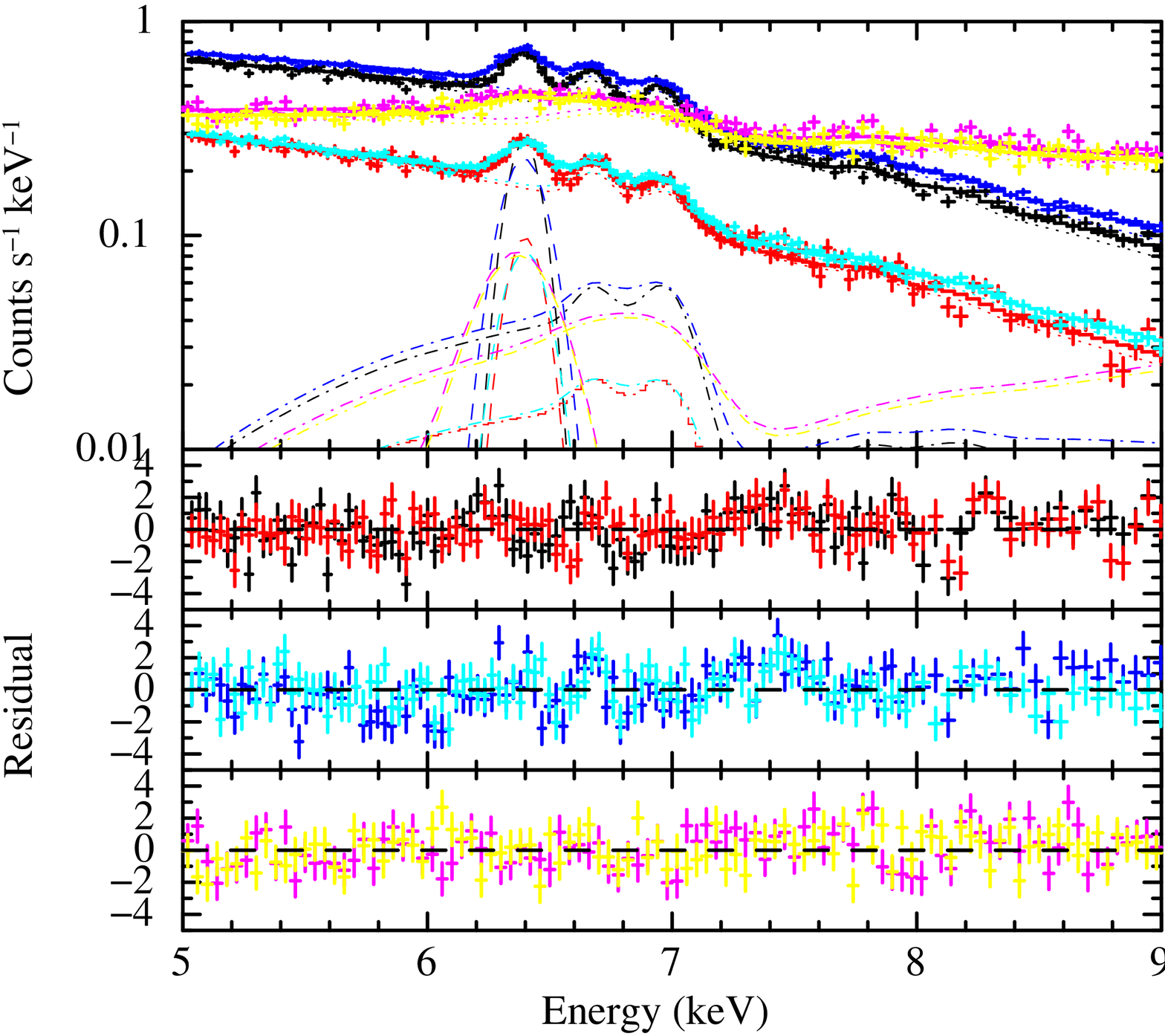}
 \caption{Same as Figure\,\ref{fig:ave_fit_acrad}, although the fitting model is 
{\tt PCA-reflect}.
 Dotted, dotted-dashed, and dashed lines show the thermal ({\sc acrad$_{\rm th}$}), fluorescent iron K$_\alpha$ line ({\sc Gaussian}), and reflection ({\sc pcfabs}) components, respectively.}\label{fig:ave_fit_pc}
\end{figure}

First, we fitted each model to the spin-averaged spectra of all datasets.
Figures\,\ref{fig:ave_fit_acrad}, \ref{fig:ave_fit_reflect}, and \ref{fig:ave_fit_pc}
display the spectrum with the best-fit models, and 
Table\,\ref{table:ave_spe_para} shows the best-fit parameters.
The three fittings are comparable in terms of the goodness of fit. 
In other words, the {\tt PCA-reflect} model reproduces the Compton hump 
as much as the {\tt IP-reflect} and {\tt AGN-reflect} models do, 
although the parameters of {\sc pcfabs} in {\tt PCA-reflect} have no physical meaning.
Here, {\tt IP-reflect} estimated the WD mass $M_{\rm WD}$ to be $0.86\pm0.01$\,$M_{\odot}$, 
which is consistent with 
$M_{\rm WD}$ = $0.87\pm0.01$\,$M_{\odot}$ and $0.83\pm{0.02}$\,$M_{\odot}$
estimated by {\tt AGN-reflect} and {\tt PCA-reflect}, respectively.
All three models estimated an extremely high specific accretion rate, with log ($a$ [\,g\,cm$^{-2}$\,s$^{-1}]) > 2$, and were found to agree with each other. 
In addition, {\tt IP-reflect} tightly constrained the viewing angle to $i = 54.2^{-2.2}_{+2.1}$$^\circ$,
which disagrees with the output of {\tt AGN-reflect} of $i < 42$$^\circ$.
The shock height $h$ and the maximum plasma temperature 
$T_{\rm max}$ (i.e., the temperature just below the shock) 
are hydrodynamically calculated \citep{2014MNRAS.438.2267H}
using the best-fit parameters, as shown in Table\,\ref{table:ave_spe_para}.
The luminosity of the thermal component within the 0.1--100\,keV band is estimated 
as 1$\times10^{34}$\,erg\,s$^{-1}$ by all three models
with the distance $D = 580\pm16$\,pc set as per
{\it Gaia} DATA RELEASE 2 (DR2; \citealt{2016A&A...595A...1G}).

\begin{table*}
\centering
 \caption{Best-fit parameters of the spectral fitting to the spin-averaged V1223\,Sgr spectra for all the data sets$^\ast$. 
 The parameters below the dashed line were hydrodynamically calculated with the best-fit parameters \citep{2014MNRAS.438.2267H}.}\label{table:ave_spe_para}
 \begin{tabular}{cccc}
 \hline
& {\sc phabs$\times$(acrad$_{\rm th}$+acrad$_{\rm ref}$}) & {\sc phabs$\times$(reflect$\times$acrad$_{\rm th}$+Gaussian)} & {\sc phabs$\times$(pcfabs$\times$acrad$_{\rm th}$+Gaussian)}\\
Nickname & {\tt IP-reflect} & {\tt AGN-reflect} & {\tt PCA-reflect} \\\hline
$N_{\rm H}$ ($\times10^{22}$\,cm$^{-2}$) & $6.5\pm0.5$ & $8.9\pm0.7$ & $8.2_{-2.5}^{+2.2}$ \\
$M_{\rm WD}$ (M$_{\odot})$ & $0.86\pm0.01$ & $0.87\pm{0.02}$ & $0.83\pm{0.02}$ \\
log($a$) (g\,cm$^{-2}$\,s$^{-1}$) & $> 2.5$ & $> 2.7$ & $> 2.0$\\
 $Z$ (Z$_\odot^\dagger$) & $0.32\pm0.01$ & $0.29\pm0.01$ & $0.26_{-0.01}^{+0.02}$ \\
 $i$ ($^\circ$) & $54.2_{-2.2}^{+2.1}$ & $< 42$ & -\\
$\Omega/2\pi$& - & $0.44_{-0.06}^{+0.10}$ & -\\
$N_{\rm H, PCA}^\ddagger$ ($\times10^{22}$\,cm$^{-2}$) & - & - & $90_{-29}^{+30}$\\
$CF_{\rm PCA}^\S$ & - & - & $0.30_{-0.05}^{+0.12}$ \\
$EW_{\rm FeI-K\alpha}^\parallel$ (eV)& - & $105_{-2}^{+15}$ & $104\pm4$ \\
$F_{\rm bol}^{\flat}$ ($\times$10$^{-10}$\,erg\,s$^{-1}$) & 2.8 & 3.2 & 3.3 \\
$L_{\rm bol}^\natural$ ($\times$10$^{34}$\,erg\,s) & 1.1 & 1.3 & 1.3\\
$f^\sharp$ & $<$ 4$\times$10$^{-5}$ & $<$ 3$\times$10$^{-5}$ & $<$ 1$\times$10$^{-4}$ \\
 $\chi^2_{\rm red}$ (d.o.f.) & 1.38 (1196) & 1.31 (1194) & 1.34 (1194)\\ 
\hdashline
$T_{\rm max}^{\star}$ (keV) & 42 & 43 & 39 \\ 
$h^\P$ ($R_{\rm WD}$) & $<$ 4$\times$10$^{-4}$ & $<$ 3$\times$10$^{-4}$ & $<$ 1$\times$10$^{-4}$ \\ 
 \hline
\multicolumn{4}{l}{$^{\ast}$ The errors indicate a 90\% statistical uncertainty.}\\
\multicolumn{4}{l}{$^\dagger$ Based on \cite{1989GeCoA..53..197A}.}\\
\multicolumn{4}{l}{$^\ddagger$ The hydrogen column density of PCA.}\\
\multicolumn{4}{l}{$^\S$ The covering fraction of PCA.}\\
\multicolumn{4}{l}{$^\parallel$ The equivalent width of the fluorescent iron K$_\alpha$ line.}\\
\multicolumn{4}{l}{$^\flat$ Unabsorbed flux in 0.1--100\,keV without the reflect component.}\\
\multicolumn{4}{l}{$^\natural$ Unabsorbed luminosity in 0.1--100\,keV without the reflect component.}\\
\multicolumn{4}{l}{$^\sharp$ The fractional accreting area.}\\
\multicolumn{4}{l}{$^\star$ The temperature just below the shock.}\\
\multicolumn{4}{l}{$^\P$ Height of the shock column from the WD surface.}\\
  \end{tabular}
\end{table*}

The change in accretion rate affects the specific accretion rate and hydrogen column density.
With this in mind, we fitted {\tt IP-reflect} again,
separating the specific accretion rate and the hydrogen column density into different datasets.
Table \ref{table:ave_spe_para_sep-obs} shows the best-fitting parameters. 
The fitting was found to improve statistically,
with the $F$-test indicating that the significance in separating the parameters is $1-2\times10^{-9}$. 
The specific accretion rate and the hydrogen column density
are consistent between 07S and 14S.
By contrast, both parameters of 14N are less than those of the other datasets, 
which is qualitatively self-consistent.
The WD mass $M_{\rm WD} = 0.92\pm0.02$\,$M_{\odot}$
is 0.06\,$M_{\odot}$ higher in mass than that 
estimated by fitting 
when binding 
the specific accretion rate and the hydrogen column density across the entire dataset. 

\begin{table*}
\centering
 \caption{Best-fitting parameters of the spectral fitting to the spin-averaged V1223\,Sgr spectrum with 
 {\tt IP-reflect}$^\ast$ $a$, $N_{\rm H}$, and normalization separated into different datasets. 
 The parameters below the dashed line were hydrodynamically calculated using the best-fitting parameters 
 \citep{2014MNRAS.438.2267H}.
 The superscript signs are the same as those in Table\,\ref{table:ave_spe_para}.
 }\label{table:ave_spe_para_sep-obs}
 \begin{tabular}{cccc}
 \hline
data set & 07S & 14S & 14N\\\hline
$N_{\rm H} (\times10^{22}$\,cm$^{-2}$) & $6.3_{-0.9}^{+0.8}$ & $6.5\pm0.5$ & $5.5_{-1.1}^{+1.2}$ \\
$M_{\rm WD}$ (M$_{\odot})$ & \multicolumn{3}{c}{$0.92\pm0.02$} \\
log($a$) (g\,cm$^{-2}$\,s$^{-1}$) & $> 1.7$ & $> 2.3$ & $0.5_{-0.2}^{+0.3}$ \\
 $Z$ (Z$_\odot^\dagger$) & \multicolumn{3}{c}{$0.34\pm0.01$} \\
 $i$ ($^\circ$) & \multicolumn{3}{c}{$53.2\pm{2.1}$} \\
$F_{\rm bol}^\flat$ ($\times$10$^{-10}$\,erg\,s$^{-1}$) & 2.9 & 2.8 & 2.8\\
$L_{\rm bol}^\natural$ ($\times$10$^{34}$\,erg\,s) & 1.2 & 1.1 & 1.1 \\
$f^\sharp$ & $<$ 2$\times$10$^{-4}$ & $<$ 6$\times$10$^{-5}$ & 4$\times$10$^{-3}$\\
$\chi^2_{\rm red}$ (d.o.f.) & \multicolumn{3}{c}{1.33 (1192)} \\
\hdashline
$T_{\rm max}^\star$ (keV) & 48 & 48 & 46 \\
$h^\P$ ($R_{\rm WD}$) & $<4\times10^{-3}$ & $<9\times10^{-4}$ & $5_{-2}^{+4}\times10^{-2}$ \\
\hline
\end{tabular}
\end{table*}

\subsection{Spin phase-resolved spectrum}

\begin{figure}
\includegraphics[width=80mm]{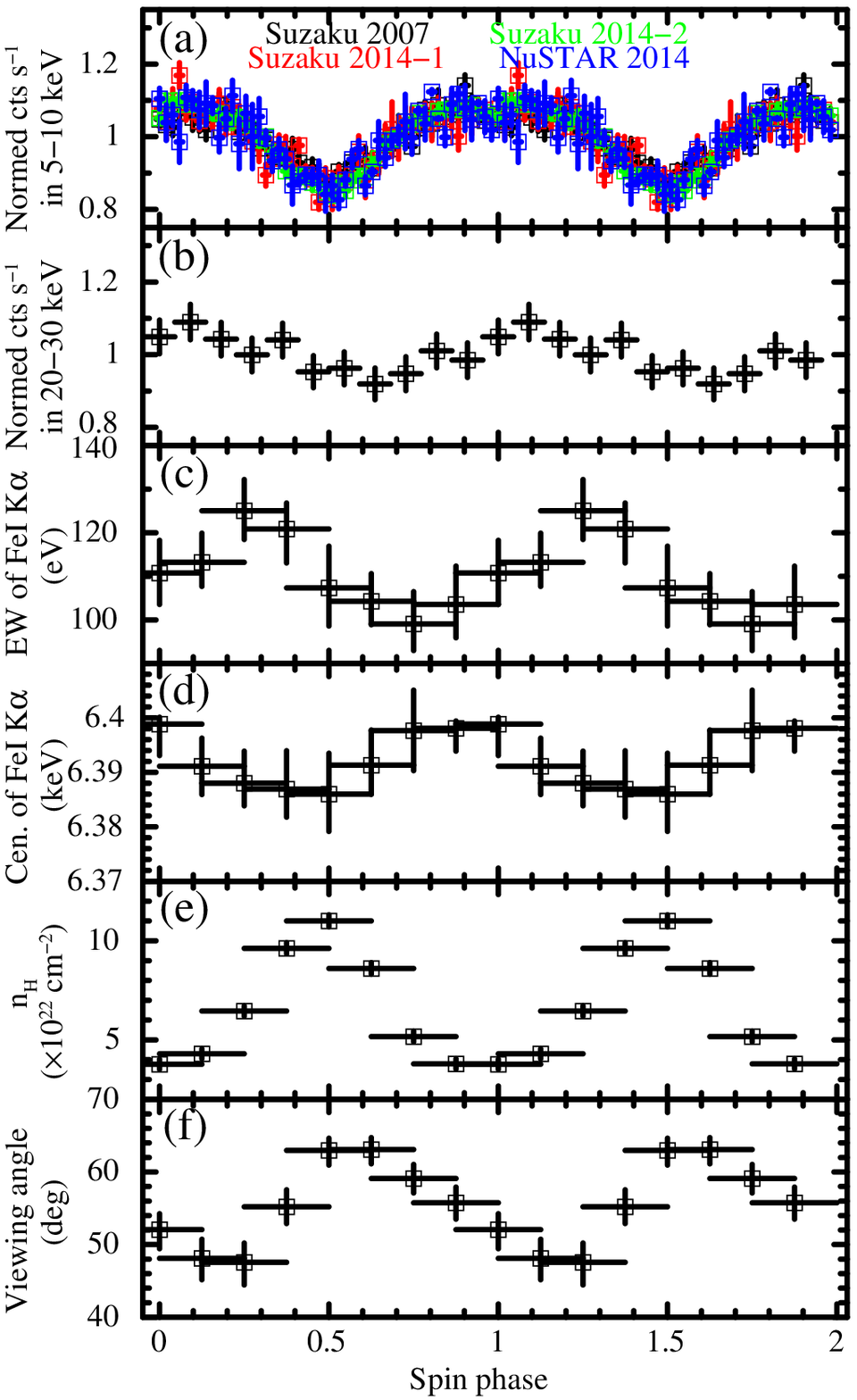}
 \caption{Modulations of several parameters in the spin period: 
 (a) 5--10\,keV folded light curves of 07S (black), 14S-March (red), 14S-April (green), and 14N (blue),
 (b) 20--30\,keV folded light curve of 14N, (c) EW of the fluorescent iron K$_\alpha$ line,
 (d) 
 Centroid of fluorescent iron K$_\alpha$ line,
 (e) Hydrogen column density of S07, 
 and (f) the viewing angle.
 The error bars in (a) and (b) denote a 68\% statistical uncertainty, and those of the other panels denote a 90\% statistical uncertainty.
(c) and (d) Outputs of the {\sc phabs$\times$(powerlaw + 3Gaussians)} fitting.
(e) and (f) Outputs of {\tt IP-reflect} fitting.
}\label{fig:para_in_sp}
\end{figure}

We divided the spectrum based on the spin phase with a period of 745.63\,s \citep{1985SSRv...40..143O}.
We determined the temporal origin of each dataset by 
 cross-correlating the light curves 
because there was no ephemeris to share among the datasets. 
We used the 5--10\,keV energy band to establish cross-correlations among the 
{\it Suzaku} and {\it NuSTAR} data. 
As a result, we determined the origins of BJD = 2454203.48031, 
2456745.50051, 2456757.50299, and 2456916.50652
for the data of 07S, 14S-March, 14S-April, and 14N, respectively.
Panel (a) of Figure\,\ref{fig:para_in_sp} shows the folded light curves.
We divided the spectrum into the following eight phases:\,0--0.25, 0.125--0.375, 0.25--0.5, 0.375--0.625, 0.5--0.75, 
0.625--0.875, 0.75--1.0, and 0.875--1.125. 
Note that these phases show a mutual overlap of 0.125. 

\begin{figure}
\includegraphics[width=80mm]{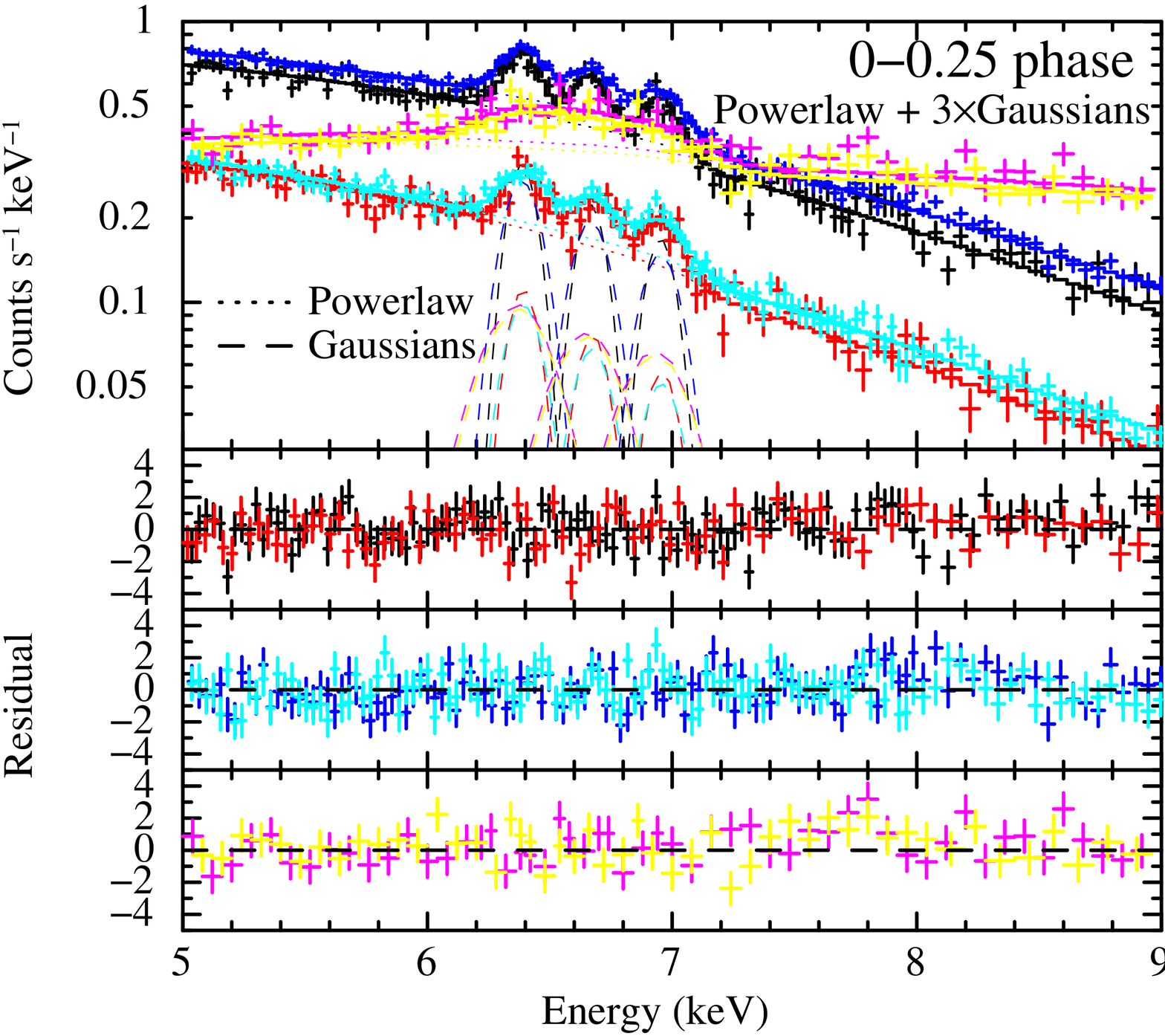}
 \caption{
The 5--9\,keV spectra from the 0--0.25 spin phase 
with the best-fitting {\sc phabs$\times$(powerlaw + 3$\times$Gaussians}) model 
to the 5--23\,keV spectra.
The relations between the data and colors are the same as in Figure\ref{fig:ave_fit_acrad}.
The dotted and dashed lines show the {\sc powerlaw} and three {\sc Gaussians}, respectively.}\label{fig:sp-res_ga}
\end{figure}

First, we fitted the {\sc phabs$\times$(powerlaw + 3$\times$Gaussian}s) model to the spin-phase-resolved spectra.
The three {\sc Gaussian}s reproduced the fluorescent, He-like, and H-like iron K$_\alpha$ lines
at 6.40, 6.70, and 6.97\,keV in the rest 
frames, respectively.
We kept the {\sc Gaussian}s narrow and used 5--23\,keV to reasonably reproduce the continuum using the power law.
Figure\,\ref{fig:sp-res_ga} shows the 0--0.25 spin phase spectrum with the best-fitting model.
This empirical model appropriately approximates the spectra.
The best-fitting energy centroid and the equivalent width (EW) 
are listed in Table\,\ref{tab:para_in_sp} along with $\chi^2$. 
Panels (c) and (d) of Figure\,\ref{fig:para_in_sp} show
the energy centroid and the EW, respectively, as functions of the phase. 
Both parameters 
modulate with the WD spin. 

Next, we fitted {\tt IP-reflect} to the spin-phase-resolved spectra.
We fixed the WD mass, abundance, and specific accretion at 
their best-fitting quantities of the phase-averaged fitting 
with the separated specific accretion rate (Table\,\ref{table:ave_spe_para}).
This is because the WD mass and abundance
should be independent of the spin phase
and the specific accretion is too insensitive to detect its spin modulation.
Moreover, we fixed the ratio of the hydrogen column densities of 
14S and 14N to that of 07S at 1.03 and 0.88, respectively, which are the best-fitting hydrogen column densities of the phase-averaged fitting. 
The consistent modulation profiles within the 5--10\,keV band across 
all datasets (panel (a) of Figure\,\ref{fig:para_in_sp})
justifies fixing the hydrogen column density ratio. 
Table\,\ref{tab:para_in_sp} shows the best-fitting parameters.
Panels (e) and (f) of Figure\,\ref{fig:para_in_sp} display
the hydrogen column density of 07S and the viewing angle, respectively, during the spin period. 
Both parameters
modulate with the WD spin as well. 

edit

\begin{table*}
\rotcaption{Best-fitting parameters of the spin-phase-resolved spectra with {\sc phabs$\times$(powerlaw + 3Gaussians)} and  {\tt IP-reflect}.}\label{tab:para_in_sp} 
\centering
\begin{sideways}
 \begin{tabular}{ccccccccc} 
\hline
 Phase & 0.875-1.125 & 0-0.25 & 0.125-0.375 & 0.25-0.5 & 0.375-0.625 & 0.5-0.75 & 0.625-0.875 & 0.75-1\\\hline
&\multicolumn{8}{c}{\sc phabs$\times$(powerlaw + 3Gaussians)} \\
FeI-K$_\alpha$ line energy (keV) & $6.391\pm0.005$ &$6.388_{-0.004}^{+0.006}$ &$6.387_{-0.005}^{+0.007}$ &$6.386\pm0.007$ &$6.391_{-0.005}^{+0.007}$ &$6.398_{-0.008}^{+0.007}$ &$6.398_{-0.004}^{+0.001}$ &$6.399_{-0.006}^{+0.001}$\\
FeI-K$_\alpha$ line $EW$ (eV) & $109.7_{-4.4}^{+4.1}$ &$120.8_{-7.8}^{+7.4}$ &$130.1_{-6.0}^{+7.2}$ &$125.9_{-5.4}^{+6.9}$ &$113.1_{-7.2}^{+5.1}$ &$109.7_{-6.6}^{+8.2}$ &$102.5_{-5.9}^{+6.3}$ &$103.6_{-7.7}^{+8.8}$\\
$\chi^2$ (d.o.f.) & 1.17 (765) & 1.08 (762) & 1.13 (753) & 1.18 (741) & 1.15 (748) & 1.12 (752) & 1.05 (759) & 1.14 (760)\\\hline
&\multicolumn{8}{c}{\sc phabs$\times$(acrad$_{\rm th}$+acrad$_{\rm ref}$)} \\
$N_{\rm H,07S}$ ($\times10^{22}$\,cm$^{-2}$) & $4.3_{-0.3}^{+0.2}$ & $6.5_{-0.3}^{+0.2}$ & $9.6\pm0.3$ & $11.0\pm0.3$ & $8.6\pm0.3$ & $5.2_{-0.3}^{+0.2}$ & $3.8_{-0.3}^{+0.2}$ & $3.8_{-0.3}^{+0.2}$\\
$N_{\rm H,14S}$ ($\times10^{22}$\,cm$^{-2}$)$^{\S}$ & 4.4 & 6.6 & 9.9 & 11.3 & 8.9 & 5.3 & 3.9 & 3.9 \\
$N_{\rm H,14N}$ ($\times10^{22}$\,cm$^{-2}$)$^{\l}$ & 3.8 & 5.7 & 8.5 & 9.7 & 7.6 & 4.6 & 3.3 & 3.3 \\
$M_{\rm WD}$ (M$_{\odot})$ & \multicolumn{8}{c}{0.92$^{\ddagger}$} \\
log($a$)$_{07S}$ (g\,cm$^{-2}$\,s$^{-1}$) & \multicolumn{8}{c}{3.4$^{\ddagger}$} \\
log($a$)$_{14S}$ (g\,cm$^{-2}$\,s$^{-1}$) & \multicolumn{8}{c}{3.3$^{\ddagger}$} \\
log($a$)$_{14N}$ (g\,cm$^{-2}$\,s$^{-1}$) & \multicolumn{8}{c}{0.5$^{\ddagger}$} \\
 $Z$ (Z$_\odot^\dagger$) & \multicolumn{8}{c}{0.34$^{\ddagger}$} \\
 $i$ ($^\circ$) & $48.1_{-2.9}^{+2.6}$ &$47.6_{-3.2}^{+2.6}$ &$55.2_{-2.3}^{+2.3}$ &$63.0_{-2.0}^{+1.6}$ &$63.1_{-2.0}^{+1.6}$ &$59.1_{-2.0}^{+1.9}$ &$55.7_{-2.2}^{+2.2}$ &$52.1_{-2.6}^{+2.2}$ \\
 $\chi^2_{\rm red}$ (d.o.f.) & 1.19 (797) & 1.12 (795) & 1.20 (785) & 1.24 (773) & 1.19 (779) & 1.18 (784) & 1.20 (791) & 1.26 (791) \\\hline
 \multicolumn{9}{l}{$^{\ast}$The errors indicate a 90\% statistical uncertainty.}\\
\multicolumn{2}{l}{$^\dagger$Based on \cite{1989GeCoA..53..197A}.}\\
\multicolumn{9}{l}{\footnotesize $^{\ddagger}$Fixed at the quantity 
best-fitted to the averaged spectrum with {\sc phabs$\times$(acrad$_{\rm th}$+acrad$_{\rm ref}$}) separating $a$ and normalization 
using the dataset.}\\
\multicolumn{9}{l}{$^{\S}$Ratio of $N_{\rm H,07S}$ was fixed at 1.02984.}\\
\multicolumn{9}{l}{$^{\l}$Ratio of	 $N_{\rm H,07S}$ was fixed at 0.88017.}\\
\end{tabular}
\end{sideways}
\end{table*}

\section{Discussion}\label{sec:dis}

\subsection{Reflection component}\label{sec:dis_ref}

We found a remarkable discrepancy in the reflection parameters 
(i.e., viewing and solid angles)
among the spin-phase-averaged spectral fittings
with the bound hydrogen column density and the specific accretion rate
(see Table\,\ref{table:ave_spe_para}).
The best-fitting viewing angles of {\tt IP-reflect} and {\tt AGN-reflect} 
are $i = 54.2_{-2.2}^{+2.1}$$^\circ$
and $< 42$$^\circ$, respectively.
The best-fitting solid angle of the {\tt AGN-reflect} model was $\Omega/2\pi = 0.44_{-0.06}^{+0.10}$.
Assuming a point source, as in {\sc reflect} in {\tt AGN-reflect}, the solid angle is expressed as
\begin{equation}
\Omega/2\pi = 1 - \sqrt{1-\frac{1}{\left(\frac{h}{R_{\rm WD}}+1\right)^2}}.
\end{equation}
Thus, the PSAC height is calculated at 13\% of the WD radius. 
Note that the PSAC should even be taller with its actual finite length. 
However, the PSAC is shorter than 0.03\% of the WD radius
according to the hydrodynamical calculation \citep{2014MNRAS.438.2267H} 
with the best-fitting parameters of {\tt AGN-reflect}, indicating that {\tt AGN-reflect} is seriously self-inconsistent.

As a major advantage of {\sc acrad$_{\rm ref}$} in {\tt IP-reflect} over other reflection models, 
it utilizes the fluorescent iron K$_\alpha$ line to determine the reflection spectrum. 
To do so, the other models use a Compton hump, which is a continuum, making it 
difficult to separate from the thermal continuum. 
In fact, the solid angle or the viewing angle is generally fixed when using {\sc reflect} (for example, \citealt{2011PASJ...63S.739H}, \citealt{2015ApJ...807L..30M}).
By contrast, we tightly constrained both parameters simultaneously using {\tt IP-reflect}.

\begin{figure}
\includegraphics[width=85mm]{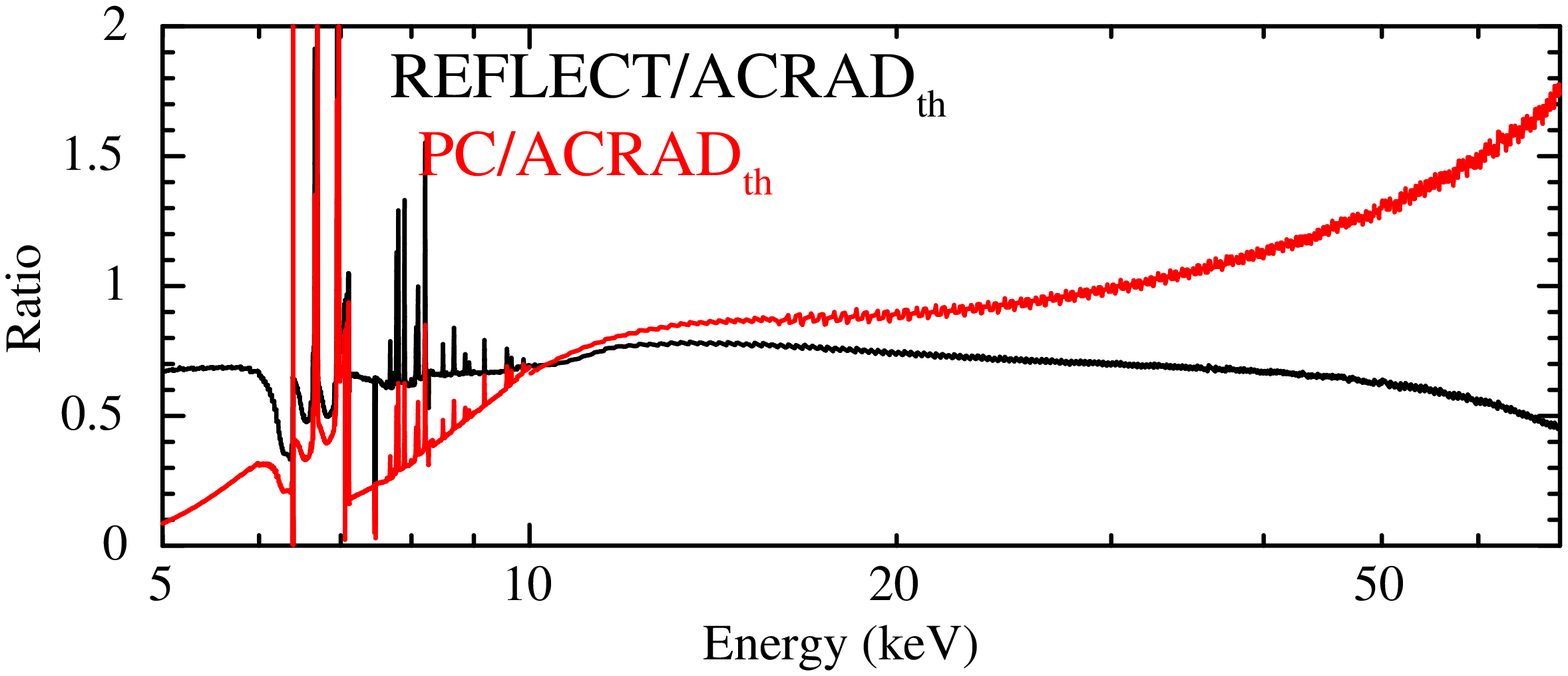}
\includegraphics[width=85mm]{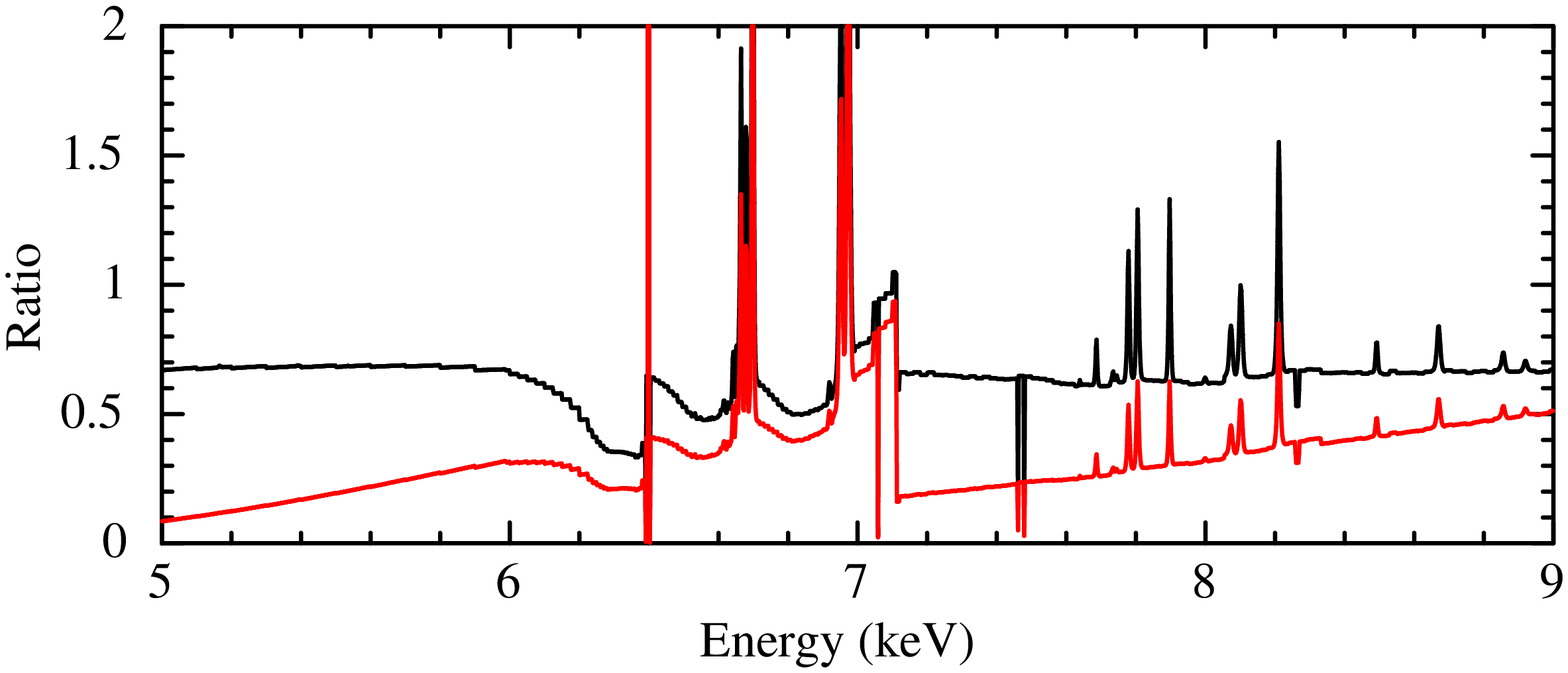}
 \caption{Spectral model ratios of the reflection component reproduced by {\sc reflect} (black) and {\sc pcfabs} (red) to that by {\sc acrad$_{\rm ref}$}.
The model spectra above and below 10 keV are the best-fitting models for FPMA and XIS-FI, respectively.
The bottom panel shows a blowup of the top at between 5 and 9keV.}\label{fig:refspe_ratio}
\end{figure}

To compare the reflection models, the ratios of the best-fit reflection spectra of 
{\tt AGN-reflect} and {\tt PCA-reflect} to that of {\tt IP-reflect} are plotted in Figure\,\ref{fig:refspe_ratio}.
In the case of {\tt PCA-reflect}, the reflection spectrum 
is the sum of the narrow Gaussian and
the thermal spectrum 
strongly attenuated using the PCA model to mimic the Compton hump.
The reflection spectrum of {\tt AGN-reflect} includes a narrow Gaussian.
The ratios have numerous lines that
originating from the thermal emission. 
{\tt IP-reflect} considers the energy loss to be due to incoherent scattering \citep{2018MNRAS.474.1810H}.
{\tt PCA-reflect} cannot reproduce the energy loss.
In addition, {\sc reflect} in {\tt AGN-reflect} does not calculate the energy loss at
below 10\,keV \citep{1995MNRAS.273..837M}.
Therefore, the 
thermal-origin emission lines in the reflection spectrum 
of {\tt PCA-reflect} and {\tt AGN-reflect} 
stay at their original energy, and the total spectrum 
overestimates the intensity of the thermal emission lines.
As a result, the reflection spectrum is inevitably suppressed
to match the overestimated model lines to the data. 
Although either an increase of the viewing angle or a decrease of the solid angle
suppresses the reflection, only the latter
results in a strong curvature at approximately 20--30\,keV 
\citep{1991MNRAS.249..352G,2018MNRAS.474.1810H}.
This stronger curvature reproduces the Compton hump better with the suppressed reflection.
As a result, the solid angle is decreased to suppress the reflection.
In fact, \cite{2015ApJ...807L..30M} reported that 
the reflection spectrum reproduced by {\sc reflect} is suppressed 
when the iron K$_\alpha$ lines are included in the fitting. 
Meanwhile, the overestimated lines also suppressed the abundance, as shown in Table \ref{table:ave_spe_para_sep-obs}.
It should be noted that a better energy resolution would
strengthen the suppression of the reflection and 
abundance
because it highlights the discrepancy in the overestimated line. 

Another noticeable feature in the spectral ratios 
is the Compton shoulder below the emission lines. 
Neither PCA-reflect nor AGN-reflect reproduces the Compton shoulder, as described above.
Moreover, {\tt IP-reflect} has fluorescent lines of neutral iron and nickel, 
unlike the PCA and {\sc reflect}.
Therefore, negative line structures are found at the energies of the 
fluorescent lines (6.404, 6.391, 7.058, 7.478, 7.461, and 8.265\,keV
for iron K$_{\alpha1,2}$, iron K$_\beta$, nickel K$_{\alpha1,2}$, and nickel K$_\beta$, respectively).

\subsection{WD mass estimation}

 In this subsection, 
 we correct the estimated WD mass by considering the finite inner disk radius described in Section \ref{sec:finit_disk_r}.
In §\ref{sec:energy_band}--\ref{sec:ref}, we discuss several factors 
that affect the WD mass measurement
and summarize the directions in Table \ref{tab:Mwd_measure_direction}.
We compare the results with an optical measurement technique in \ref{sec:opt}.

\subsubsection{Finite inner disk radius}\label{sec:finit_disk_r}
{\tt IP-reflect} assumes that accreting gas falls from infinity. 
Therefore, a non-negligible 
small inner radius results in a small shock energy release, a low shock temperature, 
and an underestimation of the WD mass 
(\citealt{2005A&A...435..191S,2019MNRAS.482.3622S}).
The inner disk radius may be approximated by the co-rotation radius,
\begin{equation}
R_{\rm co} = \left(\frac{GM_{\rm WD}P_{\rm spin}^{2}}{4\pi^2}\right)^{\frac{1}{3}}
\end{equation}
in the spin equilibrium systems \citep{1991ApJ...378..674K,1993MNRAS.261..144K,2018MNRAS.476..554S}.
With a spin period of 745.63\,s and our WD mass estimate of 0.92\,$M_{\odot}$,
the inner radius is estimated as $R_{\rm in} \sim R_{\rm co}$ = 20\,$R_{\odot}$, 
where the WD radius is computed by the relation between the WD mass and radius by \cite{1972ApJ...175..417N}.
As a result, the WD mass should be corrected by 5\% to 0.97\,$M_{\odot}$.

\subsubsection{Energy band}\label{sec:energy_band}

\begin{table*}
	\centering
	\caption{Influence of different factors on the estimation of the WD mass.}
	\label{tab:Mwd_measure_direction}
	\begin{tabular}{lp{12cm}} 	
		\hline
		Parameter & Direction\\
		\hline
		\vspace{1mm}	
		Upper limit of the energy band & It can not be too high, and the higher, the better. It should be higher than the maximum plasma temperature at least. It should be adjusted to each target by confirming that the WD mass converges as by figure\,\ref{fig:highE_Mwd}. \\
		\vspace{1mm}
		Lower limit of the energy band & It should be high enough to approximate the multicolumnar absorber by a single-column absorber but include the iron K$_\alpha$ lines. It has to be adjusted to each target by confirming that the WD mass converges as by Figure \,\ref{fig:lowE_Mwd}.\\
		\vspace{1mm}
		Reflection & It can be reproduced by the AGN reflection model ({\sc reflect}) or by the PCA model instead of the IP reflection model. Note that this substitution is not necessarily valid, with a better energy resolution than that of the CCD.\\
		\vspace{1mm}
		Abundance & It should be parameterized and free in the spectral fitting.\\
		\hline
	\end{tabular}
\end{table*}

\begin{figure}
\includegraphics[width=85mm]{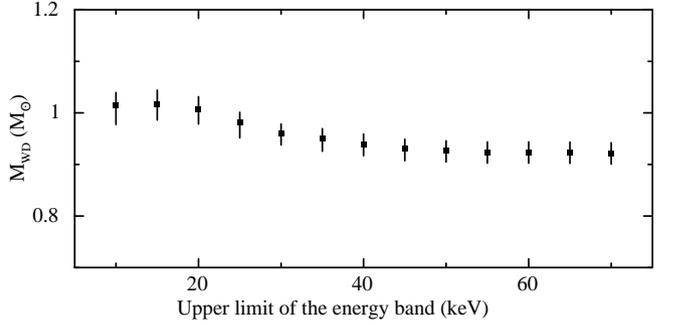}
 \caption{WD mass computed by varying the upper limit of the fitted energy band.}\label{fig:highE_Mwd}
\end{figure}

The fitting energy band is a major factor in measuring the WD mass. 
Because the maximum temperature of an IP is generally higher than 10\,keV (for example, \citealt{2010A&A...520A..25Y}),
we selected the hard X-ray spectrum \citep{2016ApJ...826..160H}. 
Figure\,\ref{fig:highE_Mwd} shows the relation between the computed WD mass 
and the upper limit of the energy band fitted with {\tt IP-reflect} 
by separating $a$, $N_{\rm H}$, and the normalization when applying the dataset. 
The computed WD mass converges on 0.92\,$M_{\odot}$ at an upper limit of approximately 50\,keV.
This is natural because the maximum temperature of the plasma in V1223\,Sgr is approximately 40\,keV.
{\it NuSTAR} is sufficiently powerful to measure the WD mass in an IP 
because it gives us high-quality 
data of up to 78\,keV. 

\begin{figure}
\includegraphics[width=85mm]{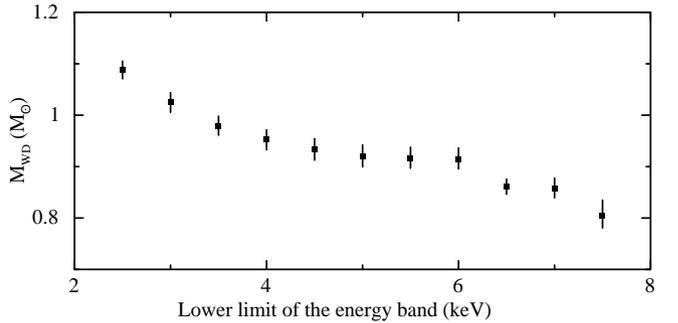}
 \caption{WD mass computed by varying the lower limit of the fitted energy band.}\label{fig:lowE_Mwd}
\end{figure}

The lower limit of the energy band is also important.
The emission lines at below 10\,keV allow us to measure the metal abundance 
and the emission measure distribution across the PSAC.
The absorption in 
an IP is created by the complicated accretion curtain and is generally difficult to model.
From an observational view, a few partial covering absorbers are needed to
reproduce the IP spectra at even below 10\,keV \citep{1999ApJS..120..277E}.
One method used to avoid this difficulty is cutting off a lower energy band \citep{1999ApJS..120..277E}.
Figure\,\ref{fig:lowE_Mwd} shows the relation between the computed WD mass and the lower limit of the energy band. 
Except for the lower limit, which is higher than 6\,keV, the WD mass converges to 0.92\,$M_{\odot}$. 
At approximately the 5\,keV lower limit, the multicolumnar absorber 
 is approximated using our applied single-column absorber.
 When the lower limit goes beyond 6\,keV (i.e., the iron K$_\alpha$ shell band), 
 the emission measure distribution in the temperature cannot
 be determined and the WD mass estimation suffers from large systematic errors.


\subsubsection{Abundance}

\begin{figure}
\includegraphics[width=85mm]{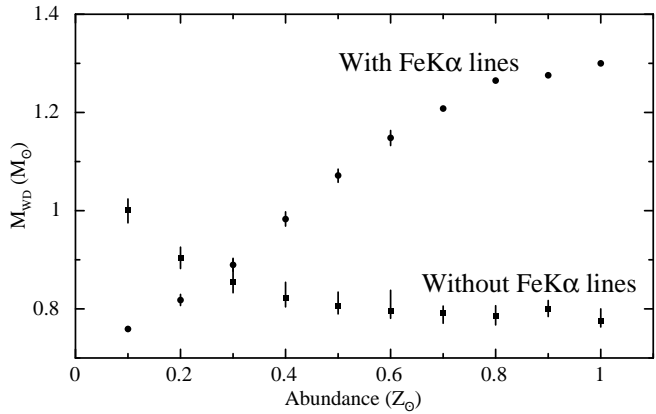}
 \caption{Computed WD mass with fixed abundance.
Squares show the data fitting without an iron K$_\alpha$ band. 
Circles show the data fitting with the iron K$_\alpha$ band. 
Errors of a part of circles are not shown (see text). }\label{fig:Z_Mwd}
\end{figure}

The abundance is another major factor in measuring the WD mass. 
In \cite{2018MNRAS.476..554S}, the authors measured the WD mass in V1223\,Sgr 
to be $0.75\pm0.02$\,$M_{\odot}$ or $0.78\pm0.01$\,$M_{\odot}$, 
which is less massive than our estimate by approximately 0.15\,$M_{\odot}$ with the {\tt IP-reflect} model.
The authors excluded the iron K$_\alpha$ band (5.5--7.5\,keV) 
and fixed the abundance at 1\,$Z_{\odot}$. 
Figure\,\ref{fig:Z_Mwd} shows the relation between the computed WD mass and the fixed abundance.
By excluding the iron K$_\alpha$ band and fixing the abundance at 1\,$Z_{\odot}$,
we obtained the WD mass of $M_{\rm WD}$ = 0.78$^{+0.02}_{-0.01}$\,$M_{\odot}$,
which agrees with \cite{2018MNRAS.476..554S}.
The lighter WD is computed with higher abundance
because the higher abundance leads to a harder reflection continuum 
{\citep{2018MNRAS.474.1810H}.
The {\tt AGN-reflect} model includes this effect as well.}
However, when we include the iron K$_\alpha$ band, 
such 
a high abundance of 1\,$Z_{\odot}$ is rejected (reduced-$\chi^2$ = 2.9).
The circles in Figure\,\ref{fig:Z_Mwd} do not have an error bar because the 
$\chi^2$ of the fittings was too large. 
With the iron K$_\alpha$ band, an overabundance leads to 
an overestimation of the WD mass 
and an underestimation of the specific accretion rate, 
and thus the emission line in the thermal and reflection spectra weakens and resembles the data.

\subsubsection{Reflection component}\label{sec:ref}

To reproduce the reflection spectrum,
we used three models, that is, {\sc acrad$_{\rm ref}$} of {\tt IP-reflect}, {\sc reflect} of {\tt AGN-reflect}, and PCA of {\tt PCA-reflect}.
The differences in the best-fitting WD mass are only within 0.04\,$M_{\rm WD}$ among the models.
This result is consistent with that of \cite{1998MNRAS.293..222C}
in that the difference in the measured WD masses 
is 0.04\,$M_{\odot}$ on average at over 13\,mCVs with and without the reflection model. 
However, as mentioned in \S\ref{sec:dis_ref}, the {\tt reflect} and PCA models 
suppress the reflection in the fitted model, and 
the degree of the suppression depends on the energy resolution.
The data that we used had a modest energy resolution, whereas the {\it Ginga} data used in 
\cite{1998MNRAS.293..222C} had an even worse energy resolution.
Therefore, we cannot judge whether reflection modeling generally influences the WD mass estimation.

\subsubsection{Comparison with optical estimation}\label{sec:opt}

 \cite{1985ApJ...289..300P} estimated the WD mass at 0.4--0.6\,$M_{\odot}$
using an optical orbital modulation measurement.
However, in this study, the maximum plasma temperature of V1223\,Sgr is estimated as 48\,keV and is no less than 30\,keV in previous studies using a multi-temperature plasma model.
As long as the Rankine-Hugoniot relations and the theoretical WD mass-radius relation are valid, the WD of 0.6\,$M_{\odot}$ leads to 22\,keV plasma even if the shock is strong and formed at the WD surface.
Therefore, the 0.4--0.6\,$M_{\odot}$ should be an underestimation. 

\subsection{Spin modulations of the parameters}

We discovered modulations of the EW of the fluorescent iron K$_\alpha$ line and viewing angle. 
The viewing angle approximately correlates with the EW and flux within the 20--30\,keV energy band. 
These anti-correlations are consistent with the fact that lowering the 
viewing angle enhances the reflection and supports the discovery regarding the viewing angle modulation.
If the reflections are a unique factor leading to modulations of the EW 
and the flux within the 20--30\,keV energy band, the modulations will correlate completely with those of the viewing angle.
However, the pre-shock gas may also contribute to the modulations, and 
thus may shift the phase of the modulations.
The fluorescent iron K$_\alpha$ lines of the EW of $\sim$ 15 and 40\,eV are 
emitted from the pre-shock gas
of $N_{\rm H}$ = 4$\times$10$^{22}$\,cm$^{-2}$ 
and 10$^{23}$\,cm$^{-2}$, respectively \citep{2011PASJ...63S.739H},
assuming that the pre-shock gas covers the plasma by 2$\pi$.
The pre-shock gas modulates the 20--30\,keV flux 
by absorption and scattering by a few percentage points. 
These effects are minor relative to the reflection but should shift the phase. 

\begin{figure*}
\begin{minipage}{0.49\hsize}
  \begin{center}
\includegraphics[width=80mm]{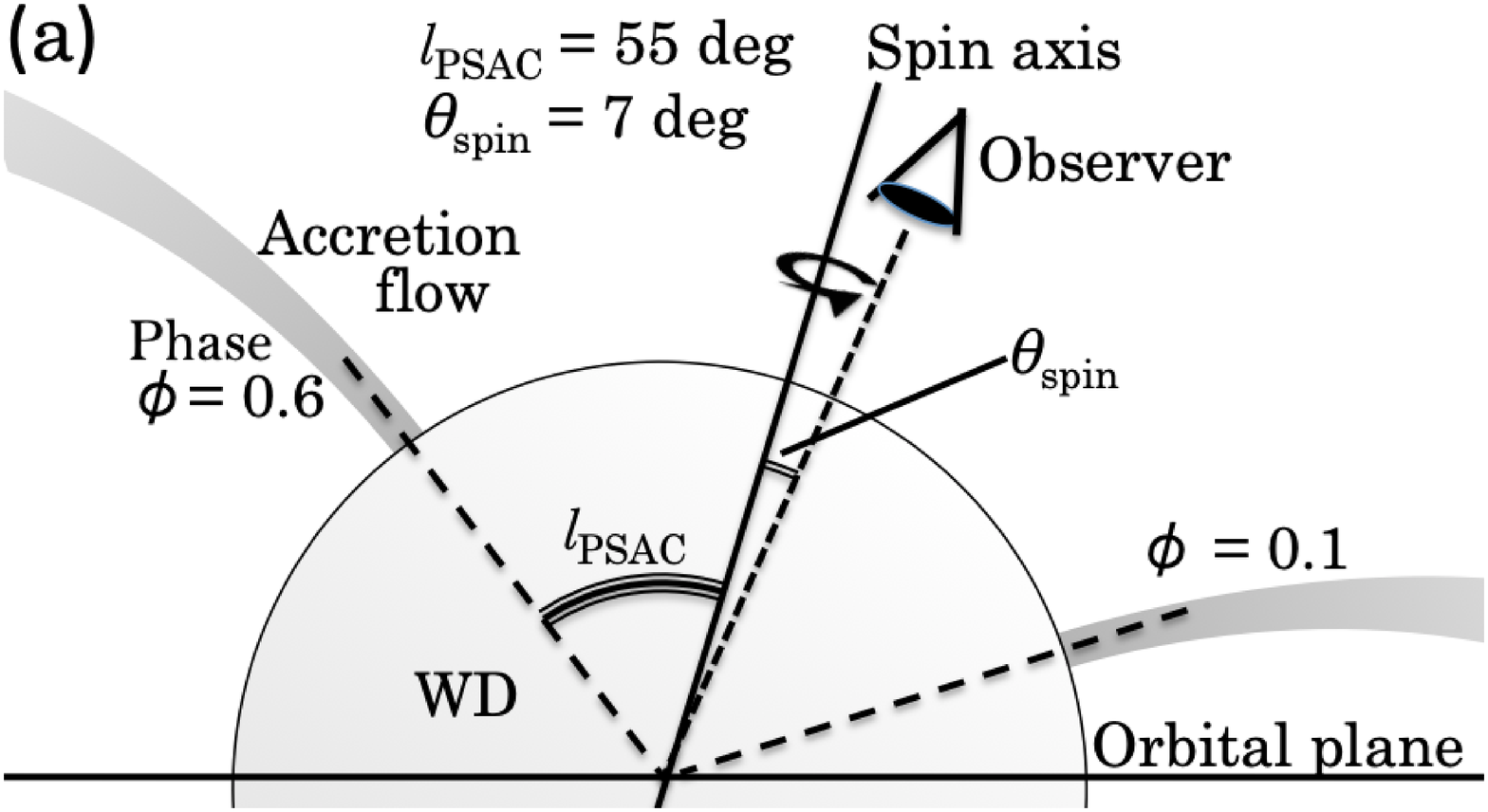}
  \end{center}
 \end{minipage}
 \begin{minipage}{0.49\hsize}
  \begin{center}
 \includegraphics[width=80mm]{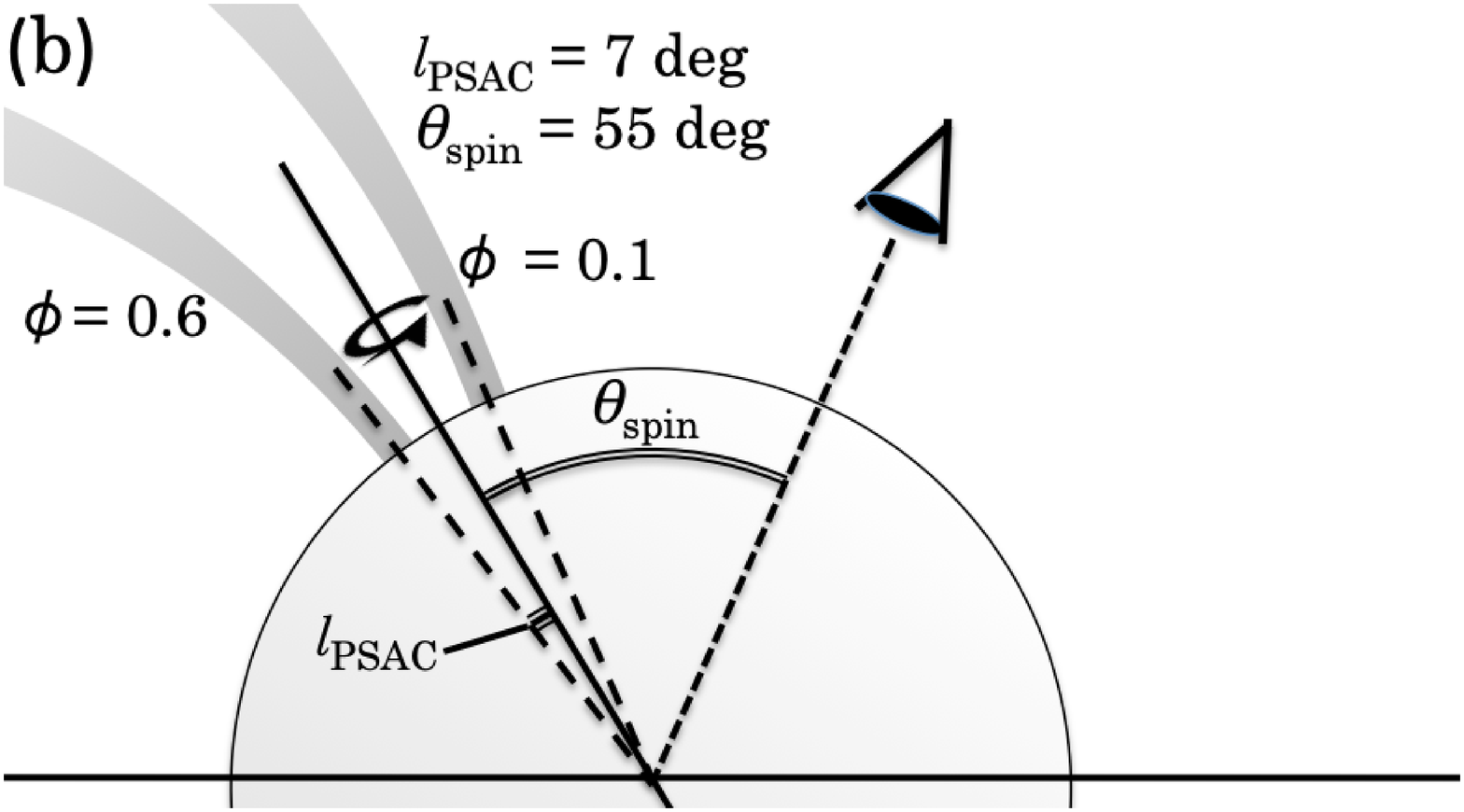}
  \end{center}
 \end{minipage}
\vspace{5mm}
 \caption{Geometries of the WD, PSAC, spin axis, and line-of-sight in V1223\,Sgr.
(a) and (b) Cases of ($l_{\rm PSAC}$, $\theta_{\rm spin}$) = (55$^\circ$, 7$^\circ$), and (7$^\circ$, 55$^\circ$), respectively.
The system inclination is 24$^\circ$
\citep{2004A&A...419..291B} in both cases.}\label{fig:geo}
\end{figure*}

A sinusoid approximates the viewing angle modulation 
with average and semi-amplitude values of 55$^\circ$ and 7$^\circ$, respectively.
Its minimum and maximum values are located at spin phases $\phi$ = 0.1 and 0.6, respectively.
Two combinations of the latitude of the PSAC ($l_{\rm PSAC}$)
and
the spin axis angle from the line-of-sight ($\theta_{\rm spin}$)
are possible, that is, ($l_{\rm PSAC}$, $\theta_{\rm spin}$) = (55$^\circ$, 7$^\circ$), and (7$^\circ$, 55$^\circ$),
as shown in Figure\,\ref{fig:geo}.
Both spin axes disagree with the reported system inclination of 17-47$^\circ$
\citep{2004A&A...419..291B}.
This disagreement implies that 
WD does not receive a substantial fraction of the angular momentum of the accreting gas. 
At this point, we cannot distinguish between the two geometries, but
we may see variations in the angle between the magnetic pole and the PSAC, and distinguish between them
if we measure $l_{\rm PSAC}$ and $\theta_{\rm spin}$ more precisely.

The maximum viewing angle phase does not correspond to that of the maximum X-ray. 
The anti-correlation between the X-ray flux and hydrogen column density agrees with the standard scenario, 
where the pre-shock gas leads to X-ray modulation by photoelectric absorption \citep{1988MNRAS.231..549R}.
This scenario expects that the X-ray flux reaches the maximum 
when the PSAC points away from the observer and 
the viewing angle is at its maximum.
However, we show that the viewing angle is close to its minimum, rather than its maximum.
Some complicated factors, such as unevenness in the density 
and/or the accretion geometry, should exceed the path-length factor.

We confirmed the central energy modulation 
of the fluorescent iron K$_\alpha$ line reported by \cite{2011PASJ...63S.739H}.
The authors suggested that the energy shift is due to the Doppler effect of the pre-shock gas accreting at $\sim$5$\times10^{3}$\,km\,s$^{-1}$.
By contrast, {\tt IP-reflect} does not require additional components around the 6.3\,keV energy band, as reported, 
and the Compton shoulder compensates this component. 
Higher-resolution data are required to distinguish between these scenarios.

\subsection{Specific accretion rate and shock height}
A fitting of the data with {\tt IP-reflect} by separating the specific accretion rate 
shows that the parameter changes by the dataset as 
log($a_{07S}$\,[g\,cm$^{-2}$\,s$^{-1}$]) $> 1.7$,
log($a_{14S}$\,[g\,cm$^{-2}$\,s$^{-1}$]) $> 2.3$,
and ($a_{14N}$\,[g\,cm$^{-2}$\,s$^{-1}$]) = $0.5_{-0.2}^{+0.3}$.
The luminosity of the three datasets is 
1.1--1.2$\times$10$^{34}$\,erg\,s$^{-1}$,
indicating that the accretion rate hardly changes. 
Therefore, the fractional accreting area ($f$, the 
ratio of the PSAC cross section to the entire WD area) should be changed to explain the 
change in the specific accretion rate.

Herein, we examine the consistency in the measured specific accretion rate, hydrogen column density, and area fraction.
With a free fall, the velocity of the pre-shock gas can be expressed as
\begin{equation}
v = \sqrt{\frac{2GM_{\rm WD}}{r}},
\end{equation}
where $r$ is the distance from the WD center. 
The dipole geometry constrains the PSAC cross section as
\begin{equation}
S = S_0\left(\frac{r}{R_{\rm WD}}\right)^{3},
\end{equation}
where $S_0$ is the PSAC cross section at the WD surface.
From the mass continuity equation, $\rho v S$ = $a S_0$ = constant,
the density is calculated as 
\begin{equation}
\rho = \frac{a R_{\rm WD}^3}{\sqrt{2GM_{\rm WD}}}r^{-\frac{5}{2}}.\label{eq:density}
\end{equation}
When the PSAC is sufficiently short,
we integrate equation\,\ref{eq:density} and obtain the mass column density as
\begin{equation}
\sigma = \frac{2 a R_{\rm WD}^3}{3\sqrt{2GM_{\rm WD}}}(R_{\rm WD}^{-\frac{2}{3}}-r^{-\frac{2}{3}}).
\end{equation}
The hydrogen column density can then be obtained as 
\begin{equation}
n_{\rm H} = \frac{\sigma}{m_{\rm H}},
\end{equation}
where $m_{\rm H}$ is the hydrogen mass. 
Assuming that the X-rays pass through a path whose length is the radius of the PSAC cross-section, 
computed from $f$ in Table\,\ref{table:ave_spe_para_sep-obs},
the specific accretion rates of the datasets 07S, 14S, and 14N ($a$ in Table\,\ref{table:ave_spe_para_sep-obs})
produce hydrogen column densities of approximately
$7.9\times10^{22}$\,cm$^{-2}$,
$17\times10^{22}$\,cm$^{-2}$, and 
$2.0\times10^{22}$\,cm$^{-2}$, respectively.
These quantities correspond to 
their respective directly measured hydrogen column density by a factor of 2, and the parameters are approximately self-consistent.

In any case, it is difficult to determine 
the specific accretion rate \citep{2014MNRAS.441.3718H}.
We will obtain new information on the specific accretion rate and the area fraction 
with a higher energy resolution from the {\it X-ray imaging and spectroscopy mission (XRISM)}
and X-ray polarization from the {\it imaging X-ray polarimeter explorer (IXPE)}.

\section{Conclusion}\label{sec:con}

We applied the newly developed IP X-ray thermal and reflection spectral models ({\sc acrad}$_{\rm th}$ + {\sc acrad}$_{\rm ref}$
called {\tt IP-reflect}) to the V1223\,Sgr combined spectrum collected from the
{\it Suzaku} satellite in 2007 and 2014, and the {\it NuSTAR} satellite in 2014.
We compared our model with an AGN reflection model ({\sc reflect} with {\sc acrad}$_{\rm th}$, called {\tt AGN-reflect}), 
and a partial covering absorption model
({\sc pcfabs} with {\sc acrad}$_{\rm th}$, called {\tt PCA-reflect}).
In this study, {\tt IP-reflect} agrees with other models for the estimated WD mass.
This result shows that the reflection modeling does not significantly influence the WD mass measurement 
in the case of V1223\,Sgr, which has a moderate energy resolution ($\Delta E \gtsim 150$\,eV).

In addition, {\tt AGN-reflect} shows a serious self-inconsistent behavior. The PSAC height determined by the best-fitting solid angle
was found to be higher than that calculated hydrodynamically
using the best-fitting WD mass and the specific accretion rate by a few orders of magnitude.
This problem is probably
due to the neglect of 
the energy loss 
by the incoherent scattering at below 10\,keV.
Data with better energy resolution may make the origin of this issue clearer.

We fitted {\tt IP-reflect} by separating 
the specific accretion rate and hydrogen column density into different data sets.
The WD mass, metal abundance, and viewing angle were estimated
to be $M_{\rm WD}$ = 0.92$\pm$0.02\,$M_{\odot}$, $Z = 0.34\pm0.01$\,$Z_{\odot}$, and $i = 53.2\pm2.1$$^\circ$, respectively.
The three datasets agree on the hydrogen column density as
$N_{{\rm H,} 07S}$ = $6.3_{-0.9}^{+0.8}\times10^{22}$\,cm$^{-2}$,
$N_{{\rm H,} 14S}$ = $6.5\pm0.5\times10^{22}$\,cm$^{-2}$, and 
$N_{{\rm H,} 14N}$ = $5.5_{-1.1}^{+1.2}\times10^{22}$\,cm$^{-2}$.
By contrast, the specific accretion rate of {\it NuSTAR} in 2014, 
log($a_{14N}$ [\,g\,cm$^{-2}$\,s$^{-1}$]) = $0.5_{-0.2}^{+0.3}$,
is lower than that of the other data 
(log($a_{07S}$ [\,g\,cm$^{-2}$\,s$^{-1}$]) $> 1.7$
and log($a_{14S}$ [\,g\,cm$^{-2}$\,s$^{-1}$]) $> {2.3}$).
These specific accretion rates constrain the PSAC height as
$h_{07S} < 4\times10^{-3}$\,R$_{\rm WD}$,
$h_{14S} < 9\times10^{-4}$\,R$_{\rm WD}$,
and $h_{14N} = 5_{-2}^{+4}\times10^{-2}$\,$R_{\rm WD}$.
If the inner disk radius is approximated by the co-rotation radius 
$R_{\rm in} \sim R_{\rm co}$ = 20\,$R_{\rm \odot}$,
the WD mass should be corrected to 0.97$\pm$0.02\,$M_{\odot}$.
The change in the specific accretion rate results in a change in
 the fractional accreting area as the thermal X-ray 
luminosity is 1.1--1.2\,$\times10^{34}$\,erg\,s$^{-1}$ across all datasets.
The directly measured hydrogen column densities are approximately consistent with 
those calculated with the specific accretion rate and fractional accreting area by a factor of 2.

The energy band and the metal abundance affect the WD mass measurement,
and the choice of incorrect value introduces large systematic errors (e.g., $\ltsimscript$\,0.2\,$M_{\odot}$ in the WD mass).
Without an energy of higher than 40 keV, it was difficult to measure the maximum temperature, 
which is essential for the WD mass measurement.
The spectrum at below 5\,keV introduces a complication of the multicolumnar absorber.
The iron K$_\alpha$ energy band is also essential 
to determine the emission measure as a function of temperature. 
Incorrect metal abundance leads to over- or under-intensity of the hard X-ray continuum and emission lines, 
resulting in a large error in the WD mass measurement. 

We fitted an empirical model composed of a power law and three Gaussians or 
{\tt IP-reflect} to the spin-phase-resolved spectra.
With {\tt IP-reflect}, the WD mass, metal abundance, 
specific accretion rate, and ratio of the hydrogen column density between the data sets
were fixed to those of the best-fit parameters of the average spectral fitting. 
We discovered for the first time the modulation of the EW and viewing angle. 
The viewing angle correlates approximately with the EW and flux within the 10--30\,keV energy band.
This fact supports the discovery of the viewing-angle modulation by {\tt IP-reflect}.

The viewing angle modulation has average and semi-amplitude values of 55$^\circ$ and 7$^\circ$, respectively.
Two combinations of the latitude of the PSAC ($l_{\rm PSAC}$) and
the spin axis angle from the line-of-sight ($\theta_{\rm spin}$) 
are possible, that is, ($l_{\rm PSAC}$, $\theta_{\rm spin}$) = (55$^\circ$, 7$^\circ$), or (7$^\circ$, 55$^\circ$). 
In either case, the spin axis disagrees with the previously reported system inclination of 24$^\circ$ 
\citep{2004A&A...419..291B}.

The anti-correlation between the viewing angle and the flux is 
inconsistent with the expectation of the standard model, which is called the accretion curtain model.
For V1223\,Sgr, a complex structure in the pre-shock gas, e.g.,
the unevenness of the density, should affect the X-ray modulation.

\section*{DATA AVAILABILITY}
The {\it Suzaku} and {\it NuSTAR} data used in this study are publicly available 
in the HEASARC archive at https://heasarc.gsfc.nasa.gov/.

\section*{ACKNOWLEDGEMENTS}

The authors are grateful to all of the Suzaku and NuSTAR project members for developing
the instruments and their software, the spacecraft operations, and the calibrations. 
We thank the anonymous referee and Dr. Yang Soong for their careful review and insightful comments, 
and Editage (www.editage.com) for English language editing. 

\end{document}